\documentclass[lettersize,journal]{IEEEtran}
\usepackage{amsmath,amsfonts}
\usepackage{algorithmic}
\usepackage{algorithm}
\usepackage{array}
\usepackage[caption=false,font=footnotesize,labelfont=rm,textfont=rm]{subfig}
\usepackage{textcomp}
\usepackage{stfloats}
\usepackage{url}
\usepackage{verbatim}
\usepackage{graphicx}
\usepackage{cite}
\usepackage{graphicx}
\usepackage{epstopdf}
\usepackage{subfloat}
\usepackage{enumerate}
\usepackage{float}
\usepackage{subfig}
\usepackage{multirow}
\usepackage{color}
\usepackage{hyperref}
\usepackage{booktabs}
\usepackage{graphicx}
\usepackage[justification=centering]{caption}
\usepackage[font=scriptsize]{caption}
\usepackage{tikz}
\usepackage{bigstrut,multirow,rotating,booktabs}
\usepackage{pifont}
\usepackage{bm}
\hypersetup{
	colorlinks=true,
	linkcolor=red,
	filecolor=blue,      
	urlcolor=blue,
	citecolor=blue,
}
\newcommand*{\circled}[1]{\lower.7ex\hbox{\tikz\draw (0pt, 0pt)%
    circle (.5em) node {\makebox[1em][c]{\small #1}};}}

\begin{document}
\title{Multi-task Just Recognizable Difference for \\ Video Coding for Machines: Database, Model, and Coding Application}
\author{Junqi Liu, Yun Zhang,~\IEEEmembership{Senior Member,~IEEE}, 
Xiaoxia Huang,~\IEEEmembership{Senior Member,~IEEE},  Long Xu,~\IEEEmembership{Member,~IEEE}, and Weisi Lin, ~\IEEEmembership{Fellow,~IEEE} 
\thanks{J. Liu, Y. Zhang, and X.2.6 Huang are with the School of Electronics and Communication Engineering, Shenzhen Campus, Sun Yat-Sen University, Shenzhen 518107, China. (Email: liujq85@mail2.sysu.edu.cn, \{zhangyun2,huangxiaoxia\}@mail.sysu.edu.cn.}
 \thanks{X. Long is with the National Astronomical Observatories, Chinese Academy of Sciences, Beijing, 100012, China. (Email:lxu@nao.cas.cn).}

\thanks{W. Lin is with the School of Computer Science and Engineering, Nanyang Technological University, Jurong West, Singapore (e-mail: wslin@ntu.edu.sg).}

}
\markboth{}
{
Lin \MakeLowercase{\textit{\textit{et al.}}}:
}


\maketitle

\begin{abstract}
Just Recognizable Difference (JRD) boosts coding efficiency for machine vision through visibility threshold modeling, but is currently limited to a single-task scenario.
To address this issue, we propose a Multi-Task JRD (MT-JRD) dataset and an Attribute-assisted MT-JRD (AMT-JRD) model for Video Coding for Machines (VCM), enhancing both prediction accuracy and coding efficiency. 
First, we construct a dataset comprising 27,264 JRD annotations from machines, supporting three representative tasks including object detection, instance segmentation, and keypoint detection.
Secondly, we propose the AMT-JRD prediction model, which integrates Generalized Feature Extraction Module (GFEM) and Specialized Feature Extraction Module (SFEM) to facilitate joint learning across multiple tasks.
Thirdly, we innovatively incorporate object attribute information into object-wise JRD prediction through the Attribute Feature Fusion Module (AFFM), which introduces prior knowledge about object size and location. This design effectively compensates for the limitations of relying solely on image features and enhances the model’s capacity to represent the perceptual mechanisms of machine vision.
Finally, we apply the AMT-JRD model to VCM, where the accurately predicted JRDs are applied to reduce the coding bit rate while preserving accuracy across multiple machine vision tasks.
Extensive experimental results demonstrate that AMT-JRD achieves precise and robust multi-task prediction with a mean absolute error of 3.781 and error variance of 5.332 across three tasks, outperforming the state-of-the-art single-task prediction model by 6.7\% and 6.3\%, respectively. 
Coding experiments further reveal that compared to the baseline VVC and JPEG, the AMT-JRD-based VCM improves an average of 3.861\% and 7.886\% Bjontegaard Delta-mean Average Precision (BD-mAP), respectively.
\end{abstract}


\section{Introduction}


\IEEEPARstart{I}{mage} and video compression serves as a fundamental infrastructure for visual communication systems. 
In the past, visual signals were primarily intended for human consumption, and compression algorithms\cite{hevc,vvc} were designed based on the characteristics of the Human Visual System (HVS). Visual masking\cite{multimodaljnd,wang2024metajnd}, frequency sensitivity\cite{jiang2022toward,jiang2024rethinking}, and other perceptual characteristic \cite{hiera_JND} were leveraged to model perceptual redundancy\cite{zhang2023survey} and optimize coding efficiency. 
Nowadays, machine intelligence is widely applied in complex real-world scenarios such as the digital retina system and autonomous driving. Video Coding for Machines (VCM)\cite{duan2020video} is proposed with the core objective of reducing bit rate while maintaining recognition accuracy. Therefore, perceptual modeling machine vision and its coding application are crucial.


As one of the most critical characteristics of the HVS, the Just Noticeable Difference (JND) \cite{JND-TMM} refers to the minimum level of distortion that can significantly affect human visual perception. To exploit this property, researchers have developed various JND models that aim to ensure the distortions introduced into images or videos remain below this perceptual threshold, thus avoiding degradation in subjective visual quality.
Liu \textit{et al.} \cite{liu2019deep} proposed to address the Picture-Wise JND (PW-JND) prediction problem by employing perceptual lossy/lossless classifiers and a sliding window refinement strategy. 
By integrating spatial and temporal features, Zhang \textit{et al.} \cite{zhang2021deep} further extended this approach to Video-Wise JND estimation.
By leveraging the correlation between pixel-level JND and PW-JND, Zhang \textit{et al.}\cite{VPJND} proposed a Visual Perception-assisted JND (VP-JND) prediction model. 
Nami \textit{et al.} \cite{10222099} proposed a Multi-Task JND (MT-JND) prediction network that incorporates a shared feature backbone and task-specific heads, enabling simultaneous prediction of multiple JND values and a visual attention map. 
Cao \textit{et al.} \cite{SGJND} proposed a cross-scale attention module to aggregate multi-layer features, providing semantic guidance of image content for JND prediction.
Nami \textit{et al.}\cite{10500870} further proposed a novel multi-task framework that leverages features from reconstructed JND-quality frames or the latent space to predict three-level JND thresholds.
However, with the growing prevalence of machine vision in real-world applications, there is an increasing demand for compression algorithms tailored to machine perception\cite{gao2021recent}. Existing JND models are typically trained on datasets constructed for the HVS, such as MCL-JCI\cite{MCL-JCI}, KonJND\cite{KonJND}, MCL-JCV\cite{MCL-JCV}, and VideoSet\cite{wang2017videoset}, which limits machine-centric applications. This mismatch highlights the need to develop perceptual models and coding algorithms exploring the characteristics of machine vision systems.

Recently, VCM has been proposed to explore the rate distortion optimization for machine vision and achieve bit rate saving for the same accuracy.  
Chen \textit{et al.} \cite{chen2023transtic} proposed instance-specific prompts and task-specific prompts for the Transformer-based encoder and decoder, transferring the base codec to Image Classification (IC), Object Detection (OD), and Instance Segmentation (IS).
Li \textit{et al.} \cite{li2024image} proposed a spatial and frequency modulation adapter, which is parameter-efficient and compatible with various neural compression architectures, facilitating IC, OD, and IS.
Fischer \textit{et al.}\cite{fisher2025latent} proposed to leverage the visual backbone to generate adaptive weights for modulating the distribution of latent variables, thereby preserving components that are more critical to IS. 
Lu \textit{et al.}\cite{Lu2024pre} proposed a quantization-adaptive neural preprocessing module and a proxy network, eliminating machine vision redundancy before traditional encoders.
Yang \textit{et al.}\cite{Pre-Processor} proposed a lightweight task-switchable pre-processor, which suppressed irrelevant information of multiple machine vision tasks by corresponding task-attentive semantic modulation modules.
Yang \textit{et al.} \cite{Yang2024compact} proposed a codebook hyperprior design and realized compact visual representation compression for multi-task intelligent applications.
Zhang \textit{et al.}\cite{allinone} proposed the multi-path aggregation technique to construct an all-in-one compression framework, achieving a balance between task-specific adaptation and multi-task generalization.
Liu \textit{et al.}\cite{liu2024rate} proposed a flexible Rate-Distortion-Cognition controlling method within a single compressor, enabling scalable image coding for both human and machine vision.
These methods optimized VCM via tuning techniques, neural pre-processing modules, and innovative framework designs.
However, the implicit feature representations within end-to-end frameworks offer limited insight into the underlying machine vision mechanisms. 
More explicit and interpretable perceptual characteristics of machine vision remain to be explored.

Investigating the JND model for machines, recently termed Just Recognizable Difference (JRD), has drawn much attention due to its promising applications in VCM.
JRD represents the minimal perceptual threshold that significantly influences machine vision performance among various distortion types and levels.
Jin \textit{et al.}\cite{jin2021just} regarded JRD as additive noise that can be injected into a clean image while still maintaining relatively high machine vision performance.
They leveraged magnitude control and spatial constraint mechanisms to identify redundant pixel-wise features.
Zhang \textit{et al.} \cite{zhang2022smr} proposed a Satisfied Machine Ratio (SMR) prediction and constructed a database for IC and OD based on HEVC. Then, JRD was determined as SMR is larger than a given threshold. 
Liu \textit{et al.}\cite{JRD_DCC} proposed a no-reference JRD prediction for OD, which extracts key features from both the source image and the residual from source and GAN\cite{GAN}-generated JRD images. Approximately 50\% bit rate reduction was achieved when applying this JRD to image compression for YOLOv7\cite{yolov7}.

Predicting JRD can be modeled as a multi-class classification problem \cite{zhang2021just,zhang2023learning,liu2024dtjrd}.
Zhang \textit{et al.}\cite{zhang2021just} proposed an ensemble learning-based JRD prediction by leveraging multiple binary classifiers, which requires additional distorted reference images and incurs substantial model complexity.
Zhang \textit{et al.}\cite{zhang2023learning} proposed an Object-Wise JRD (OW-JRD) model for Versatile Video Coding (VVC)\cite{vvc} compressed images and OD task, where multi-class JRD prediction is decomposed as multiple binary classifications and solved by reusing a single binary classifier multiple times.
These methods require image compression and JRD prediction multiple times, causing additional computational complexity. 
To improve JRD prediction accuracy and reduce complexity, Liu \textit{et al.}\cite{liu2024dtjrd} proposed a deep ViT-based JRD framework, named DT-JRD, where Gaussian Distribution-based Soft Labels (GDSL) were used to increase the number of JRD labels. However, existing JRD research is limited to IC and/or OD and does not consider multiple machine vision tasks. In addition, existing JRD datasets and models are developed for single-task prediction and lack mechanisms for joint optimization across multiple vision tasks.
If directly using a single-task JRD model multiple times to predict Multi-Task JRD (MT-JRD), it not only reduces generalization but also requires additional computational costs due to repeated inference for each task.

To address these issues, we built an MT-JRD database and proposed an Attribute-assisted MT-JRD (AMT-JRD) model for VCM. The key contributions are summarized as follows.
\begin{enumerate}
\item  We construct an MT-JRD dataset for OD, IS, and Key Point Detection (KPD). In-depth analysis is conducted on the commonalities and differences in JRD distributions among the three tasks.
\item  We propose an AMT-JRD prediction framework, which integrates Generalized Feature Extraction Module (GFEM) and Specialized Feature Extraction Module (SFEM) for multi-task learning. Beyond that, we innovatively design an Attribute Feature Fusion Module (AFFM) to fuse the attribute features of object size and location with image features to enhance the modeling capability of object-wise JRD.
\item  We propose an MT-JRD-based VCM using VVC and JPEG as the base codecs. Experiments show that compared to the VVC and JPEG baseline, the AMT-JRD-based VCM achieves an average of 3.861\% and 7.886\% BD-mAP gains, respectively.
\end{enumerate}

\begin{figure}[t]
\captionsetup{justification=justified}
    \centering
        \subfloat[]{
        \includegraphics[width=0.38\textwidth]{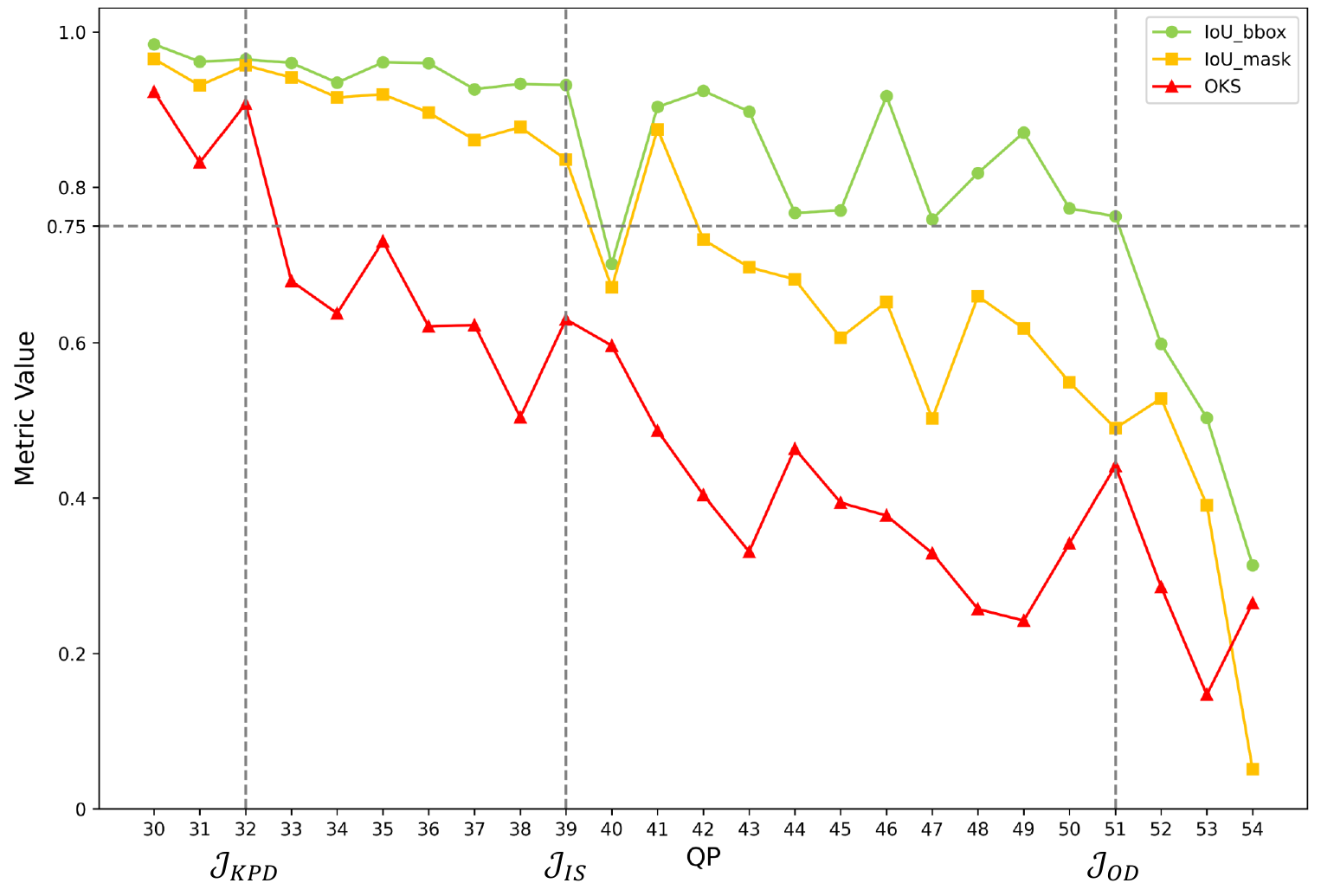}}
        \\
        \subfloat[original]{
        \includegraphics[width=0.15\textwidth]{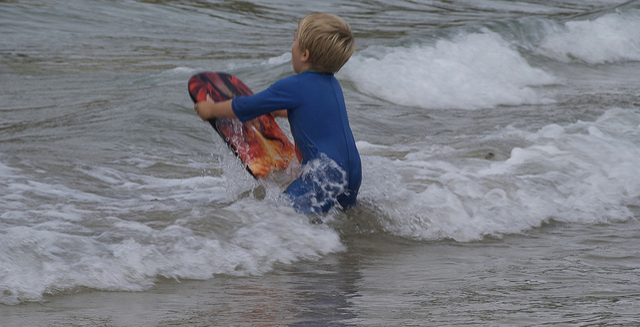}}
        \hspace{0.05em} 
        \subfloat[]{
        \includegraphics[width=0.15\textwidth]{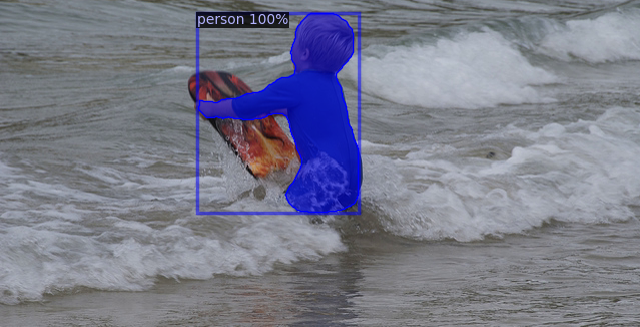}}
        \hspace{0.05em} 
        \subfloat[]{
        \includegraphics[width=0.15\textwidth]{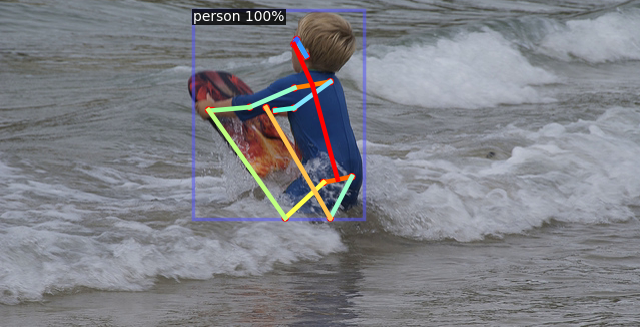}}
        \\[-2ex]
        \subfloat[$J_{KPD}=32$]{
        \includegraphics[width=0.15\textwidth]{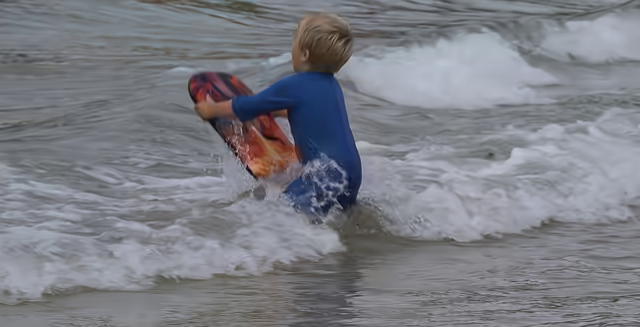}}
        \hspace{0.05em} 
        \subfloat[]{
        \includegraphics[width=0.15\textwidth]{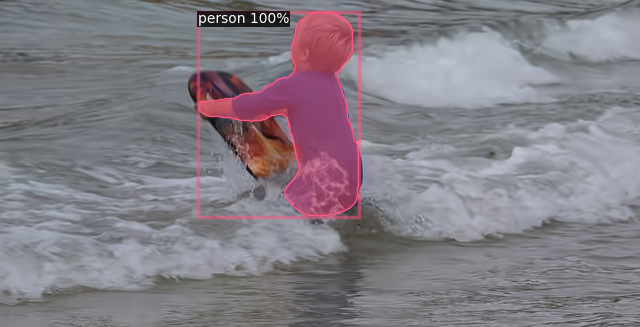}}
        \hspace{0.05em} 
        \subfloat[]{
        \includegraphics[width=0.15\textwidth]{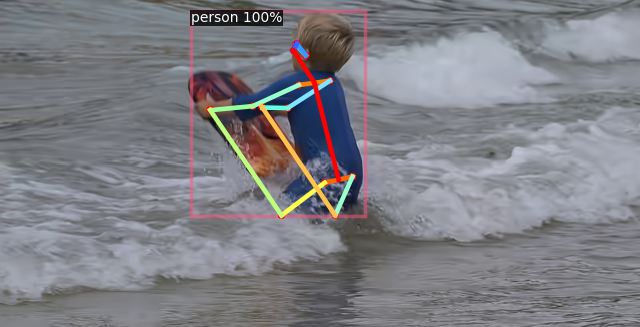}}
        \\[-2ex]
        \subfloat[$J_{IS}=39$]{
        \includegraphics[width=0.15\textwidth]{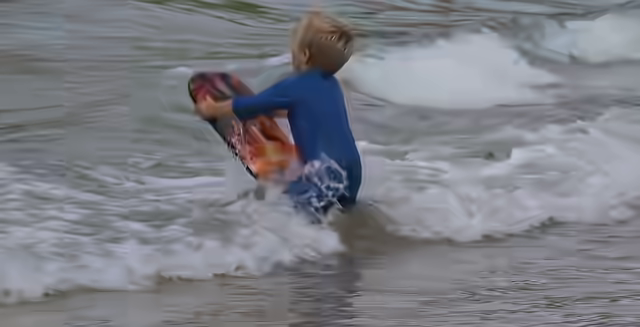}}
        \hspace{0.05em} 
        \subfloat[]{
        \includegraphics[width=0.15\textwidth]{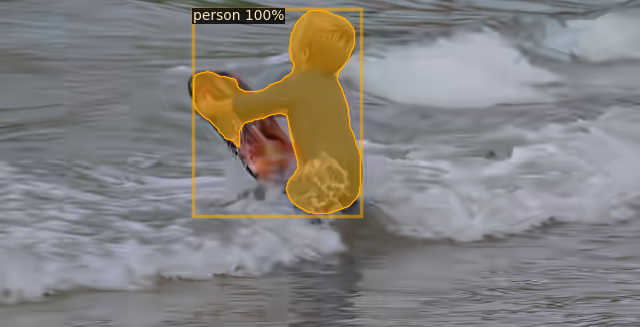}}
        \hspace{0.05em} 
        \subfloat[]{
        \includegraphics[width=0.15\textwidth]{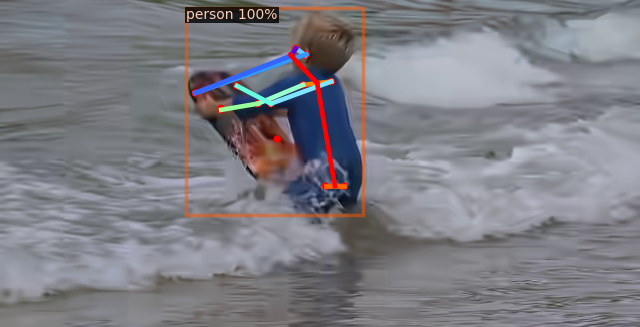}}
        \\[-2ex]
        \subfloat[$J_{OD}=51$]{
        \includegraphics[width=0.15\textwidth]{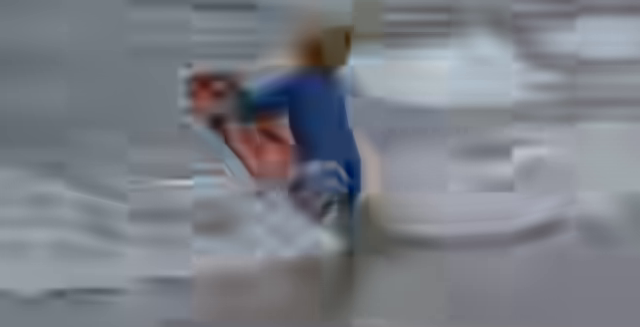}}
        \hspace{0.05em} 
        \subfloat[]{
        \includegraphics[width=0.15\textwidth]{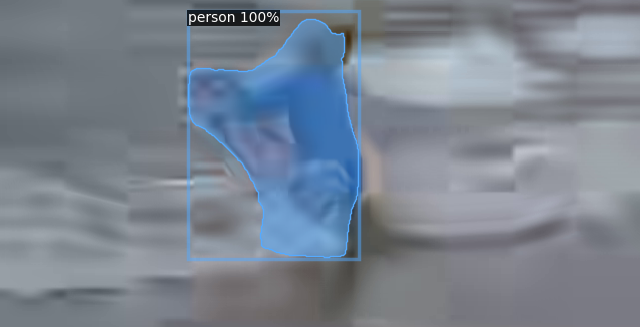}}
        \hspace{0.05em} 
        \subfloat[]{
        \includegraphics[width=0.15\textwidth]{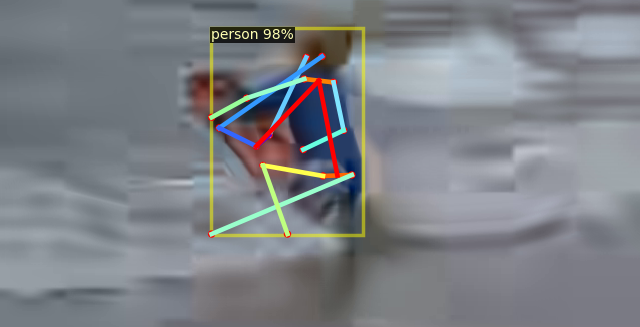}}
    \caption{Visualized examples of MT-JRD. (a) Performance degradation in multi-task scenarios. In subfigures (b)-(m), the first column displays the original image or the distorted image compressed with MT-JRDs; the second and third columns present the corresponding OD/IS and KPD results.}
    \label{example_22360}
\end{figure}

\begin{figure}[t]
	\centering
        \captionsetup{justification=justified}
        \includegraphics[width=0.48\textwidth]{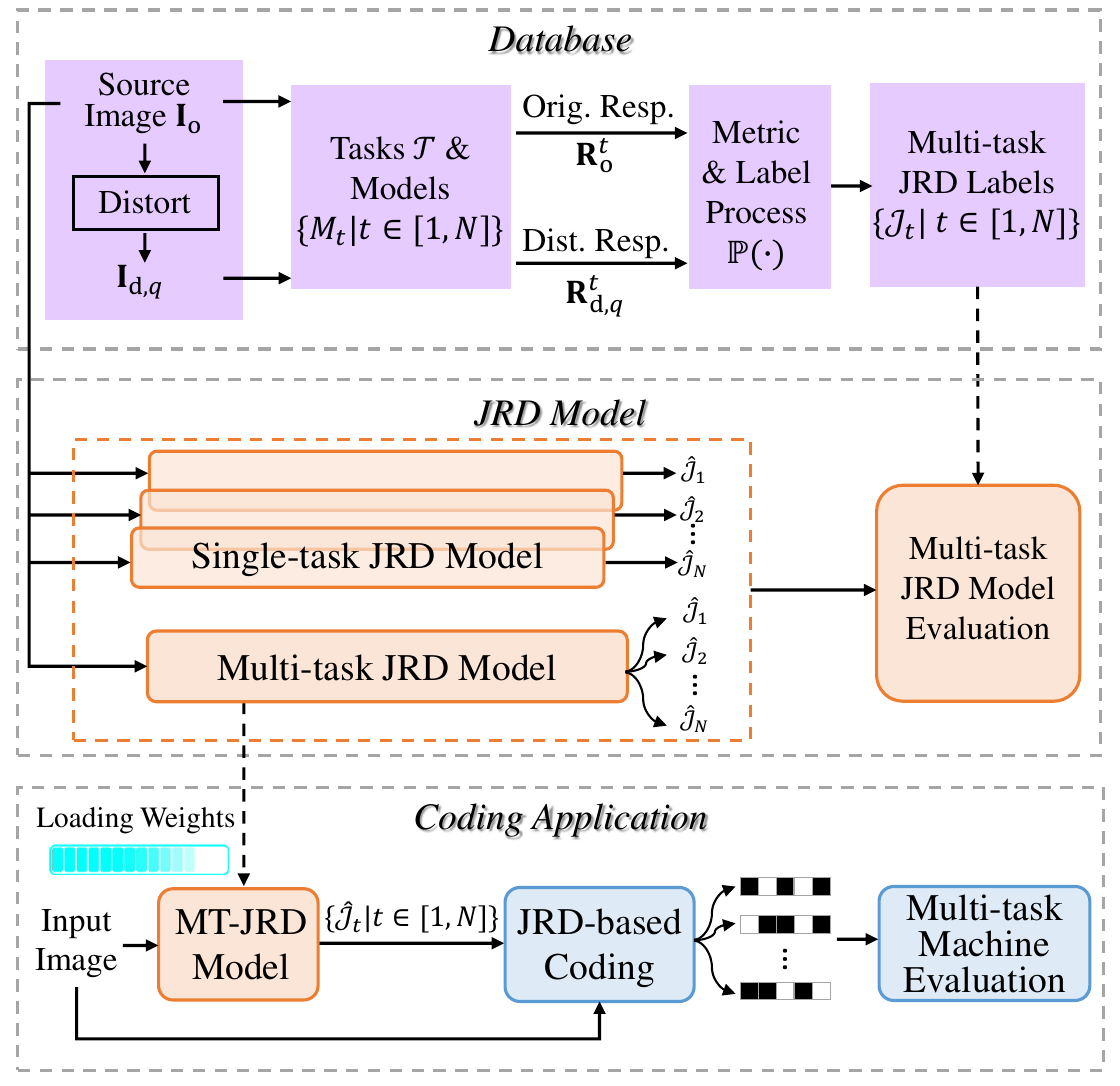}
        \caption{Paradigm of MT-JRD including database, model, and coding application.}
        \label{paradigm}
\end{figure}

The remainder of this paper is organized as follows. Section \ref{section:motivation} describes the motivation and problem formulation of MT-JRD for machines. Section \ref{section:dataset} describes the detailed pipeline of MT-JRD dataset construction. Section \ref{section:framework} presents the framework and key modules of the AMT-JRD prediction model. Section \ref{section:vcm} introduces the JRD-based image coding for machines. Section \ref{section:experiment} presents the experimental results and analysis. Section \ref{section:conclusion} draws the conclusion.

\section{Motivation and Problem Formulation}
\label{section:motivation}
In practice, multiple different vision tasks, such as OD\cite{fasterrcnn} and IS\cite{maskrcnn}, are usually performed for a compressed image.
Due to different features and semantic visual properties, there is a substantial gap between JRDs across different machine vision tasks.
To analyze the JRD gap in multi-task scenarios, we compressed images with VVC and all 64 quality levels \cite{zhang2023learning}, and then evaluated their accuracy degradations in three representative tasks, i.e., KPD, IS, and OD.
Fig. \ref{example_22360}(a) shows the accuracy degradation for VVC compressed images with different Quantization Parameters (QPs). We can observe the accuracies decline differently across tasks, where the Object Keypoint Similarity (OKS)\cite{coco} for KPD degrades fastest, and the Intersection over Union (IoU) scores for IS and OD degrade in the second and third places. If we set a threshold with 0.75\cite{zhang2023learning}, the $J_{KPD}$, $J_{IS}$, and $J_{OD}$ are 32, 39 and 51, respectively. They validate the different JRD properties in machine vision. 
Moreover, it indicates that KPD/IS is more sensitive to the compression distortion than OD. Fig. \ref{example_22360}(b)-(m) show the visualized examples of the MT-JRDs. The JRD for one task, e.g., OD, may cause recognizable distortion for the other, e.g., KPD. Also, JRD for KPD will be too strict for IS/OD. 
Due to the different properties of JRD, investigating the MT-JRD model is highly demanded.

In modelling MT-JRD, one straightforward solution is to decompose multiple tasks and predict JRD $\hat{J}_{t}$ for task $t$ individually, which is
\begin{equation}
    \hat{J}_{t} = f_{\mathrm{ST}}^{t}(\mathbf{I}_{\mathrm{o}})
    \label{singletaskmodel}
\end{equation}
where $\mathbf{I}_{\mathrm{o}}$ is the source image, $f_{\mathrm{ST}}^{t}$ denotes the JRD prediction model for task $t$, $t\in\{1,2,\dots,N\}$ and $N$ is the number of tasks. Multiple JRD models $f_{\mathrm{ST}}^{t}$ shall be learned and run multiple times to predict all $N$ JRDs for $N$ tasks, which is straightforward, but inefficient, time-consuming, and architecturally complicated.

To address these problems, one MT-JRD model $f_{\mathrm{MT}}$ that predicts JRDs for multiple machine vision tasks in one pass is desired, which can be formulated as 
\begin{equation}
   \{\hat{J}_{t}\}_{t=1}^{N}=f_{\mathrm{MT}}(\mathbf{I}_{\mathrm{o}})
\end{equation}
The $f_{\mathrm{MT}}$ is capable of exploiting common features among multiple tasks to achieve high accuracy, low inference complexity, and a compact model architecture.
Fig. \ref{paradigm} shows the paradigm of MT-JRD and its applications for VCM, including the MT-JRD database, the prediction model, and the coding application.

\section{MT-JRD Database and Analysis}
\label{section:dataset}
\begin{figure*}[t]
	\centering
        \captionsetup{justification=justified}
        \includegraphics[width=0.95\textwidth]{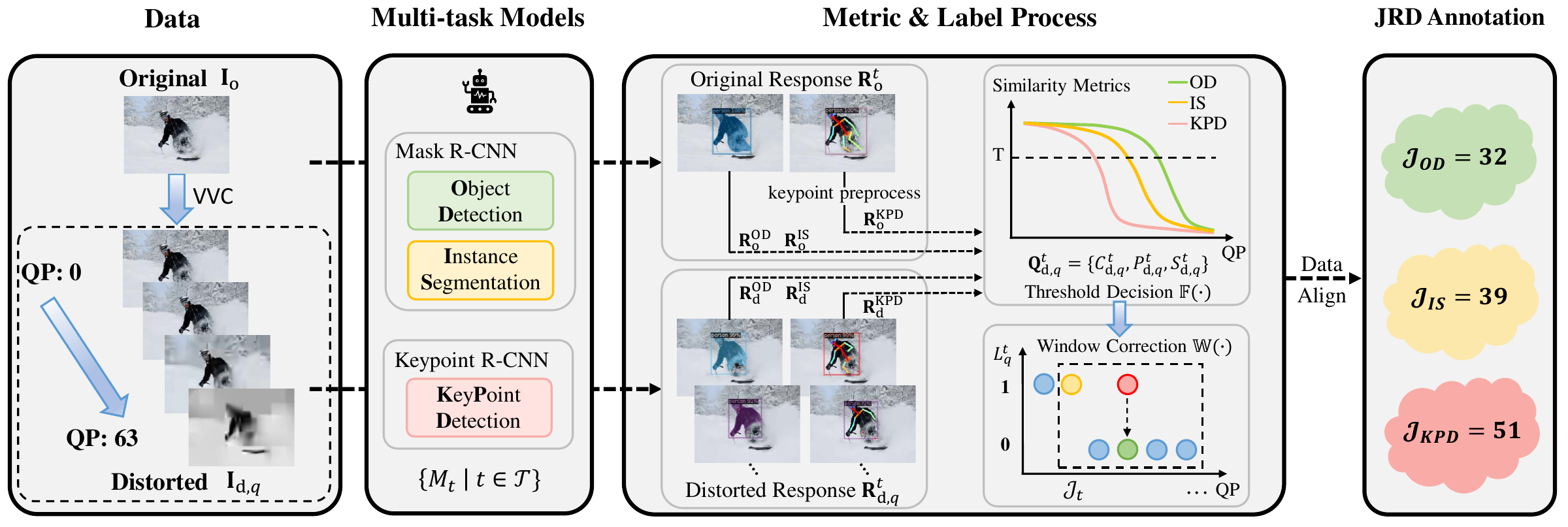}
        \caption{The pipeline for constructing the MT-JRD dataset.}
        \label{fig2}
\end{figure*}

\subsection{MT-JRD Database Establishment}
The MT-JRD dataset is built on a subset of the COCO\cite{coco} testing set, which contains diverse scenes and supports a wide range of machine vision tasks.
We primarily investigate the impact of compression distortion on the multi-task machine vision.
Each source image $\mathbf{I}_{\mathrm{o}}$ is compressed by VVC with 64 QPs to generate a set of 64 distorted versions $\mathcal{I}_{\mathrm{d}}=\{\mathbf{I}_{\mathrm{d},q}\mid q \in \{0,1,\dots,63\}\}$.
We construct the MT-JRD dataset based on three representative machine vision tasks, namely OD, IS, and KPD. They perform different levels of machine vision semantic information and are recommended by the MPEG VCM group. 
Mask R-CNN\cite{maskrcnn} is employed for both OD and IS owing to its high detection and segmentation accuracy, while Keypoint R-CNN is used for KPD.
Both models use ResNeXt-101\cite{resnext101} as the backbone, and their pretrained weights are available on the Detectron2 platform.

Fig. \ref{fig2} shows the pipeline for constructing the MT-JRD dataset. First, we feed the original and each distorted image, i.e., $\mathbf{I}_{\mathrm{o}}$ and $\mathbf{I}_{\mathrm{d},q}$, into the machine vision model $M_t$ to obtain the original response and distorted response at distortion level $q$ for task $t$ as
\begin{equation}
\left\{
\begin{aligned}
    \mathbf{R}_{\mathrm{o}}^{t} &= M_t(\mathbf{I}_{\mathrm{o}})= \{C_{\mathrm{o}}^t, P_{\mathrm{o}}^t, O_{\mathrm{o}}^t\} \\
    \mathbf{R}_{\mathrm{d},q}^{t} &= M_t(\mathbf{I}_{\mathrm{d},q})=\{C_{\mathrm{d},q}^t, P_{\mathrm{d},q}^t, O_{\mathrm{d},q}^t\} 
\end{aligned}
\right.
\end{equation}
where $C$ and $P$ denote the class and confidence probability, respectively. $O$ represents task-specific output, i.e., the bounding box for OD, the mask for IS, and the keypoint vector for KPD. Keypoint preprocessing is performed on $\mathbf{R}_{\mathrm{o}}^{\mathrm{KPD}}$, including the approximation of the segmentation mask area, object size filtering, and visibility label assignment, to enable the standard OKS calculation \cite{coco}.

Second, similarity measurement for task $t$, denoted as $S_{\mathrm{d},q}^t$, is calculated between $O_{\mathrm{o}}^t$ and $O_{\mathrm{d},q}^t$. For OD and IS tasks, we employ $\text{IoU}_{\phi}$\cite{pascalvoc} as $S_{\mathrm{d},q}^t$ metric, where $\phi \in \{B, M\}$ denote the bounding box and mask, respectively. For the KPD task, we utilize OKS\cite{coco} to present $S_{\mathrm{d},q}^t$, which measures the similarity between keypoint vectors.
Third, an individual label $L_q^t$ is obtained by mapping the machine quality $\mathbf{Q}_{\mathrm{d},q}^t = \{C_{\mathrm{d},q}^t, P_{\mathrm{d},q}^t, S_{\mathrm{d},q}^t\}$ through a thresholding function $\mathbb{F}(\cdot)$ as
\begin{equation}
L_q^t = \mathbb{F}(\mathbf{Q}_{\mathrm{d},q}^t)=
\begin{cases}
1, & (C_{\mathrm{d},q}^t = C_{\mathrm{o}}^t) \land (P_{\mathrm{d},q}^t > T) \land (S_{\mathrm{d},q}^t > T) \\
0, & \text{otherwise}
\end{cases}
\end{equation}
where the threshold $T$ balances the recognizable ratio and distortion. $T$ is set as 0.75 \cite{zhang2023learning}.
After obtaining $L_q^t$, a sliding window correction strategy \cite{zhang2023learning} is applied to identify the transition point from unrecognizable to recognizable distortion, which defines the JRD. Ultimately, we constructed JRD annotations from machines, comprising 9,088 person-class instances per task from 4,348 images.

\subsection{Statistical Analysis on the MT-JRD dataset}

\begin{figure}[t]
	\centering
        \captionsetup{justification=justified}
        \subfloat[]{
        \includegraphics[width=0.24\textwidth]{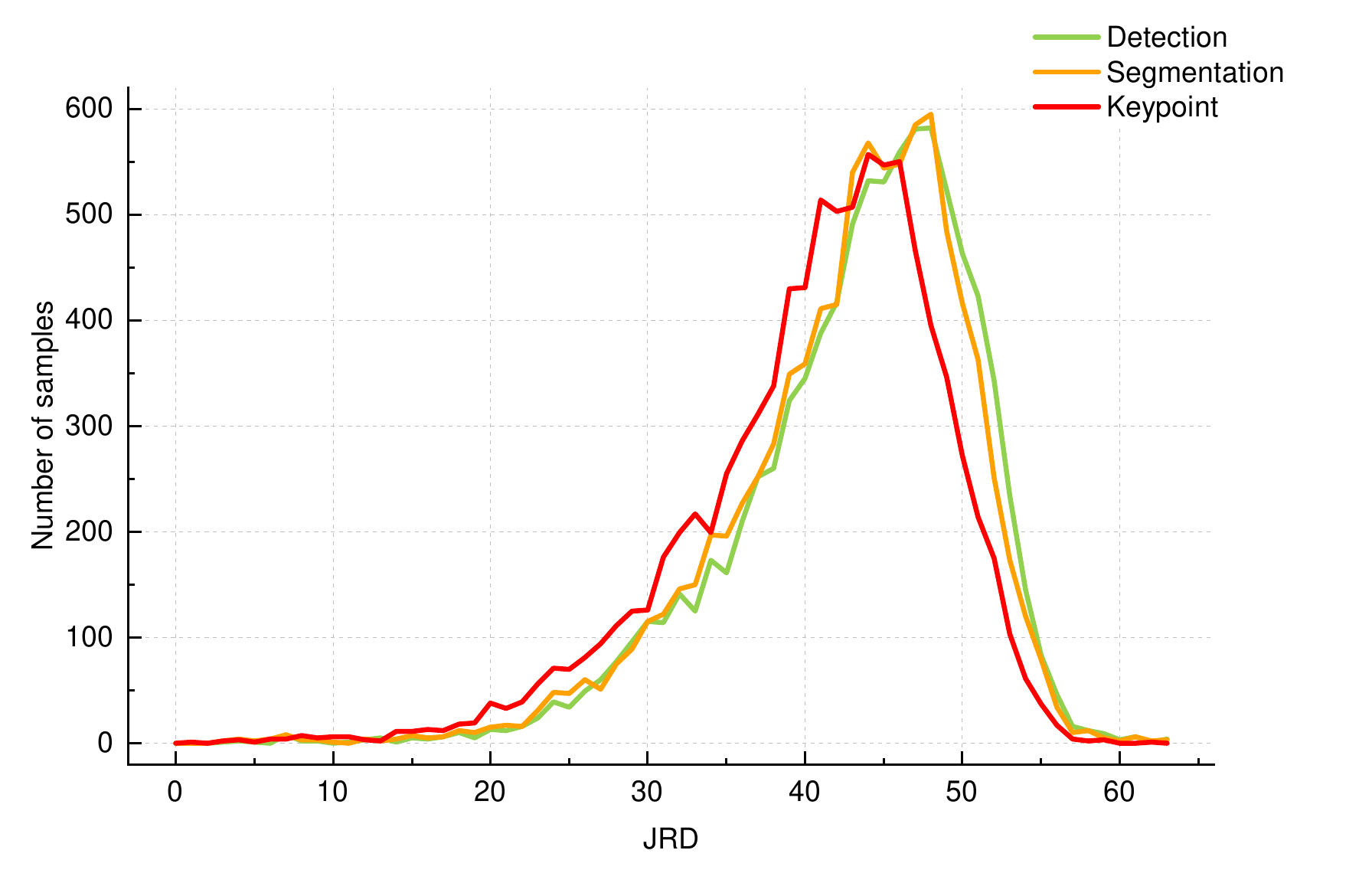}}
        \subfloat[]{
        \includegraphics[width=0.24\textwidth]{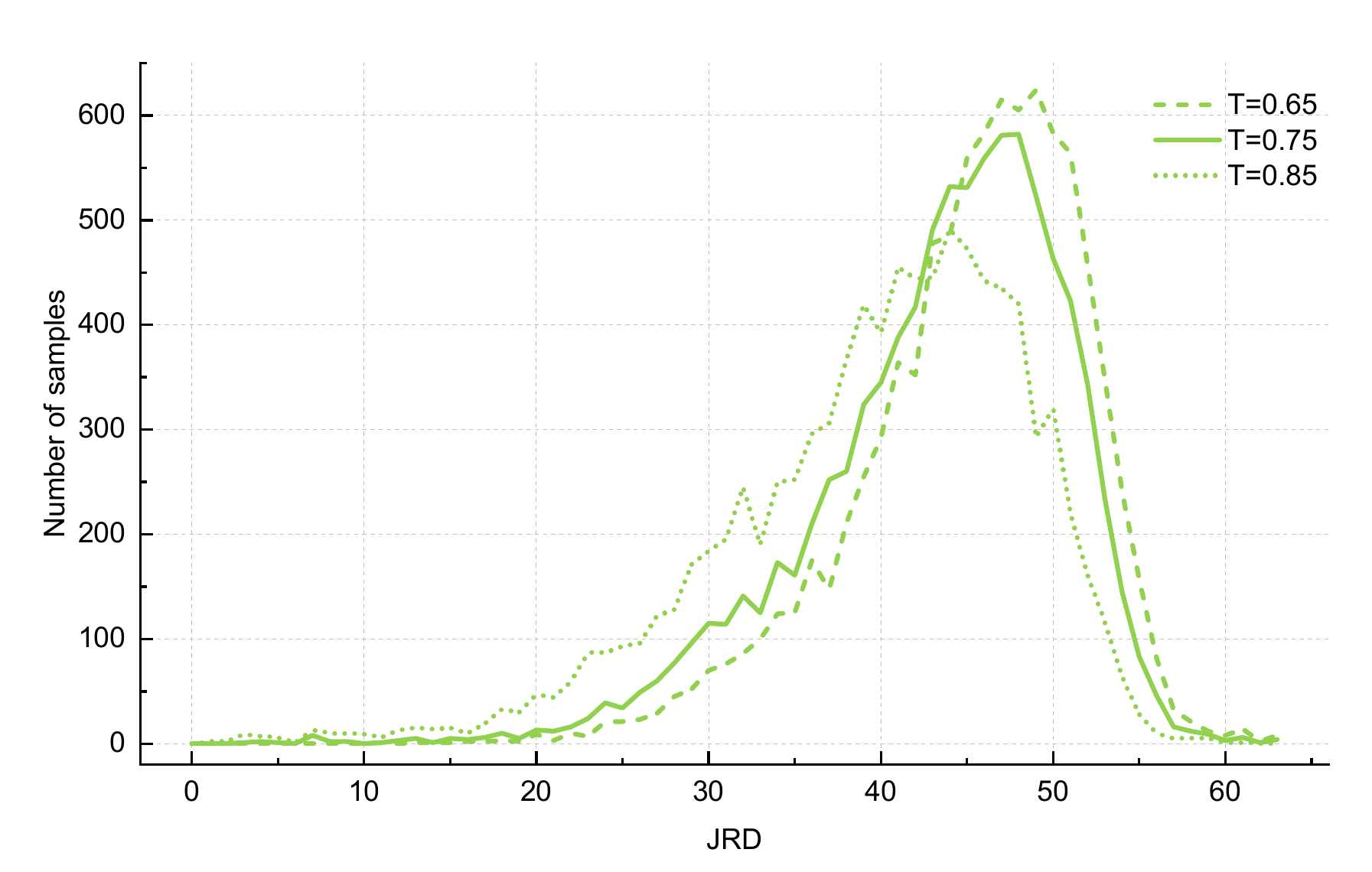}}
        \\[-2ex]
        \subfloat[]{
        \includegraphics[width=0.24\textwidth]{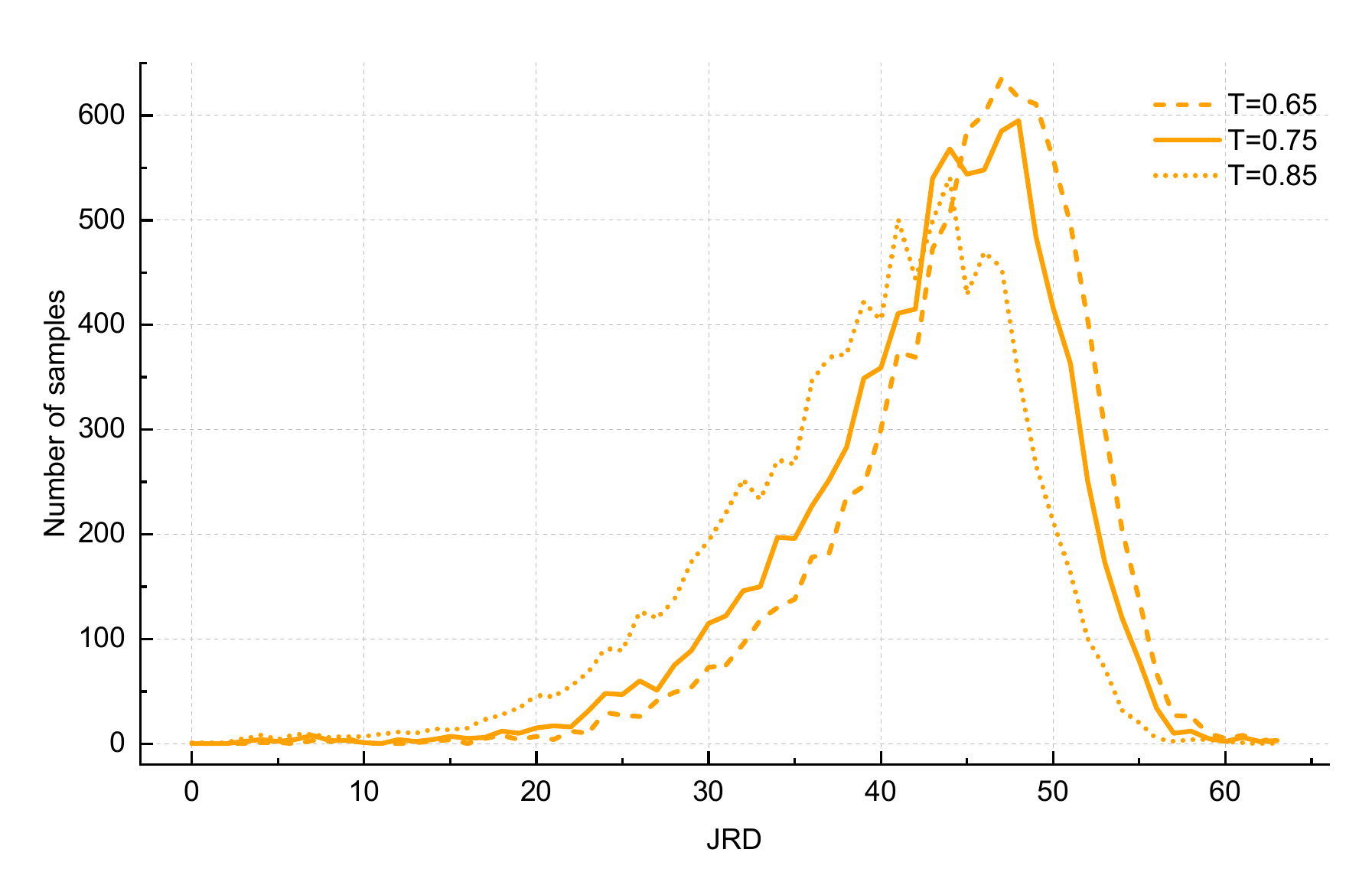}}
        \subfloat[]{
        \includegraphics[width=0.24\textwidth]{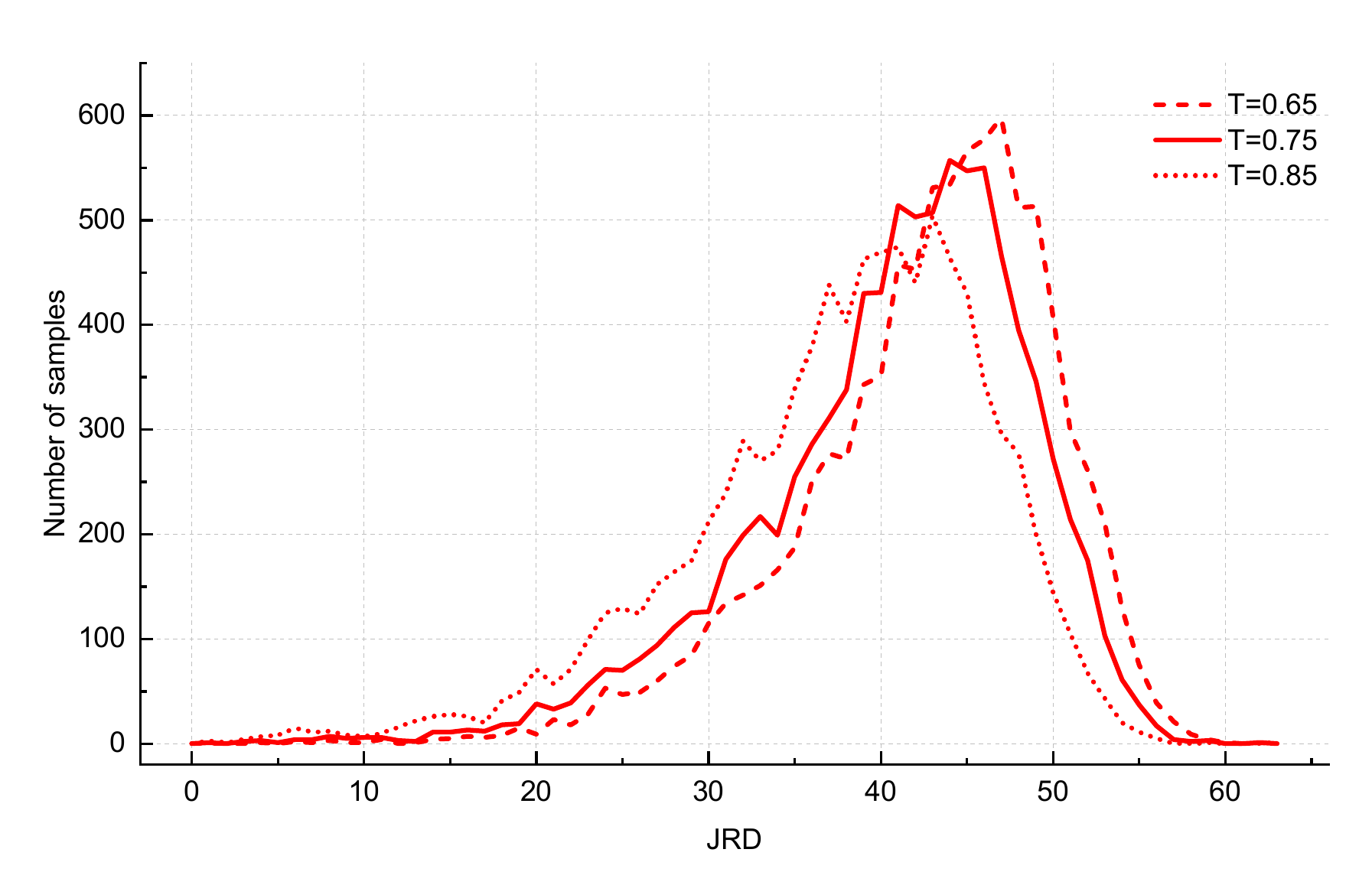}}
        \caption{MT-JRD quantity distribution under different $T$s. (a) MT-JRD distribution when $T=0.75$, (b) OD JRD, (c) IS JRD, (d) KPD JRD.}
        \label{JRD_distribution}
\end{figure}

\begin{figure*}[t]
    \centering
        \subfloat[]{
        \includegraphics[width=0.3\textwidth]{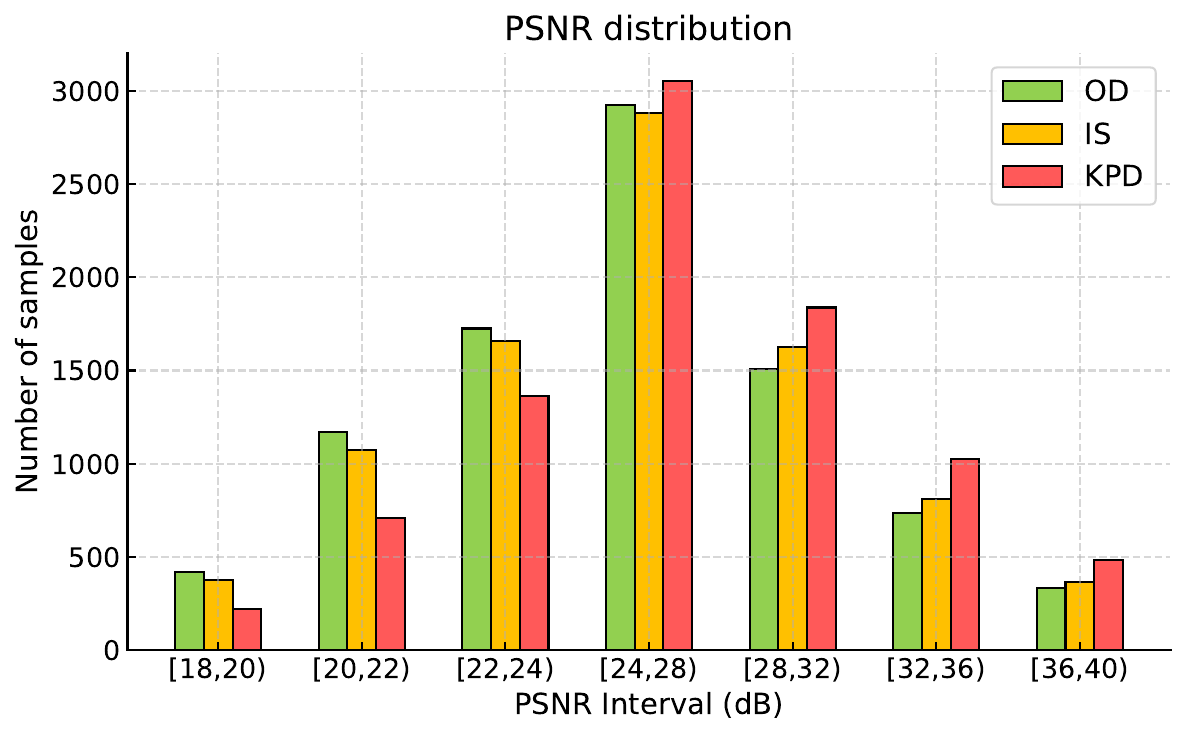}}
        \subfloat[]{
        \includegraphics[width=0.3\textwidth]{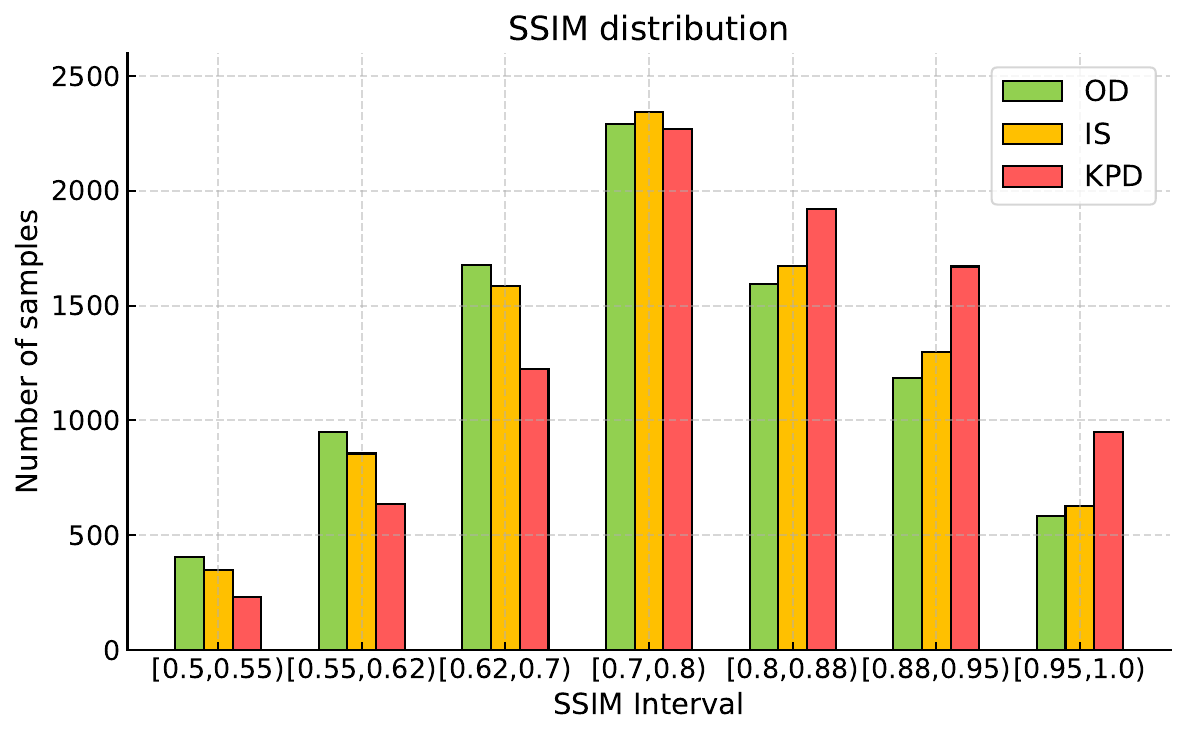}}
        \subfloat[]{
        \includegraphics[width=0.3\textwidth]{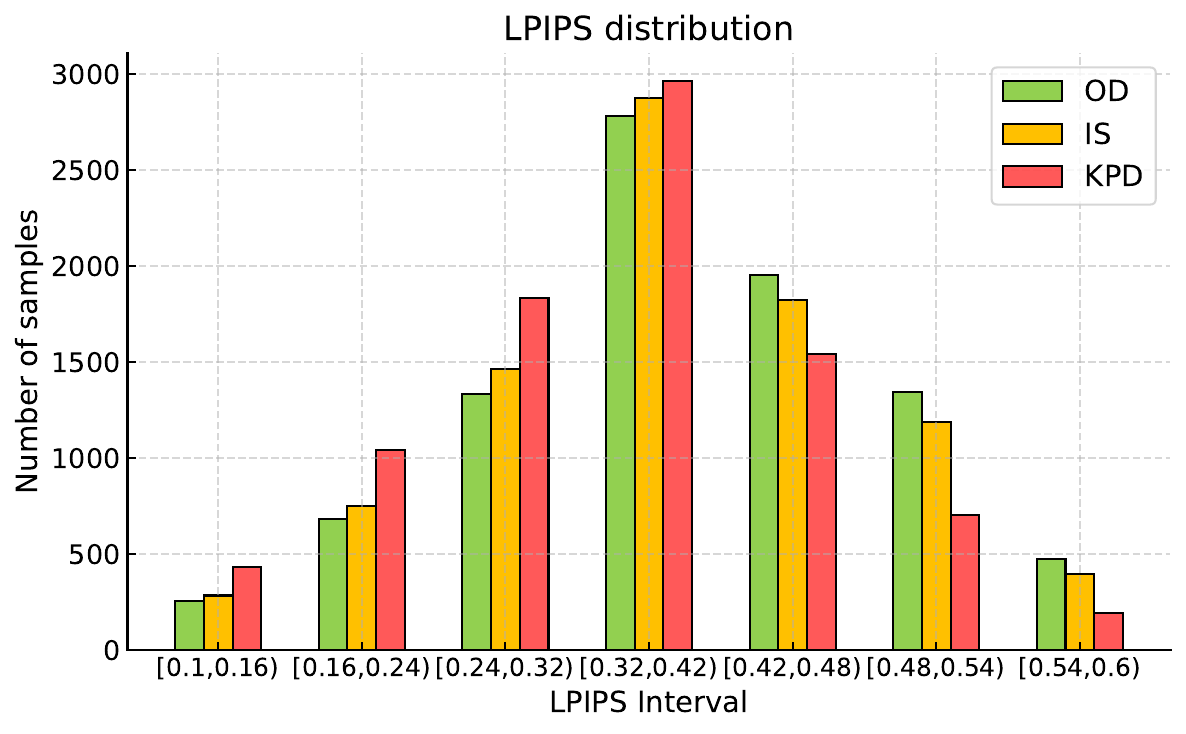}}
        \caption{Distortion distribution of the MT-JRD dataset.}
	\label{quality distribution}
\end{figure*}

\subsubsection{Similarity and Difference of MT-JRD}
First, as illustrated in Fig. \ref{JRD_distribution}, the JRD values across the three tasks demonstrate a broadly similar distribution pattern. Specifically, more than 85\% of the samples for each task are concentrated within the JRD range spanning from 27 to 51, indicating that perceptual degradation predominantly occurs within a common sensitivity band. 
Moreover, the distributions are asymmetric, with a larger proportion of samples on the lower-JRD side of the peak, suggesting that as quantization-induced distortion increases exponentially, progressively fewer samples can maintain reliable recognition under high distortion levels.
Overall, these observations reflect a partially aligned perceptual sensitivity mechanism across tasks.

Second, despite this general similarity, the three tasks exhibit discernible variations in their JRD distributions, reflecting their different perceptual sensitivities.
As shown in Fig.~\ref{JRD_distribution}(a), the red curve representing KPD is clearly left-shifted relative to the curves for IS and OD.
Quantitatively, the average JRD for OD, IS, and KPD is 43.296, 42.680, and 40.658, respectively. 
As shown in Fig. \ref{quality distribution}, the OD samples are predominantly concentrated in the lower PSNR/SSIM\cite{SSIM} and higher LPIPS\cite{LPIPS} intervals. Conversely, KPD samples are mostly distributed in the higher PSNR/SSIM and lower LPIPS intervals.
It is found that KPD exhibits the lowest tolerance to quantization-induced distortion, IS shows a moderate level of tolerance, and OD is the most distortion-tolerant. 
A plausible explanation lies in the intrinsic differences in task requirements. While OD primarily relies on coarse object localization through bounding boxes, both IS and KPD demand finer-grained outputs such as pixel-level segmentation masks or spatially precise keypoint coordinates and their structured connectivity. 
These dense predictions are inherently more sensitive to spatial and structural degradations, leading to a significant drop in recognition under equivalent distortion conditions.

\begin{figure}[t]
    \centering
        \includegraphics[width=0.4\textwidth]{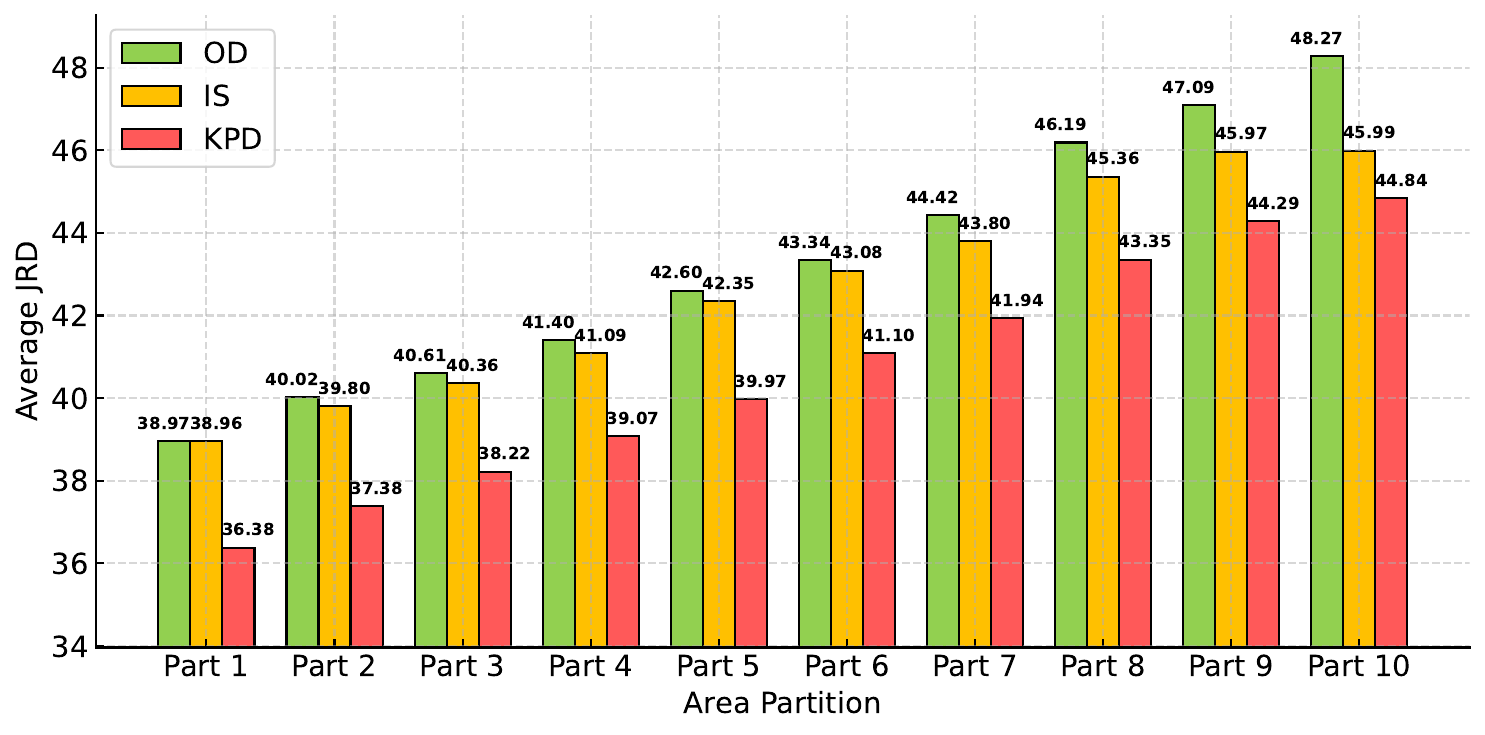}
        \caption{Relationship between object size and JRD.}
	\label{MTJRD size}
\end{figure}
\begin{figure}[t]
    \centering
        \subfloat[]{
        \includegraphics[width=0.16\textwidth]{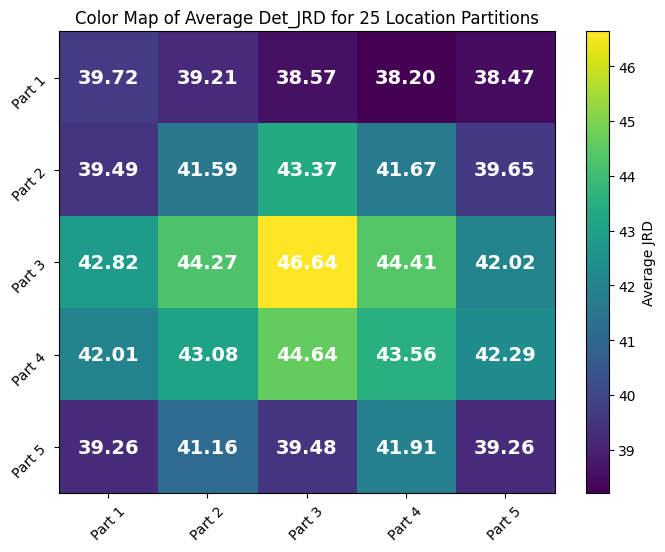}}
        \subfloat[]{
        \includegraphics[width=0.16\textwidth]{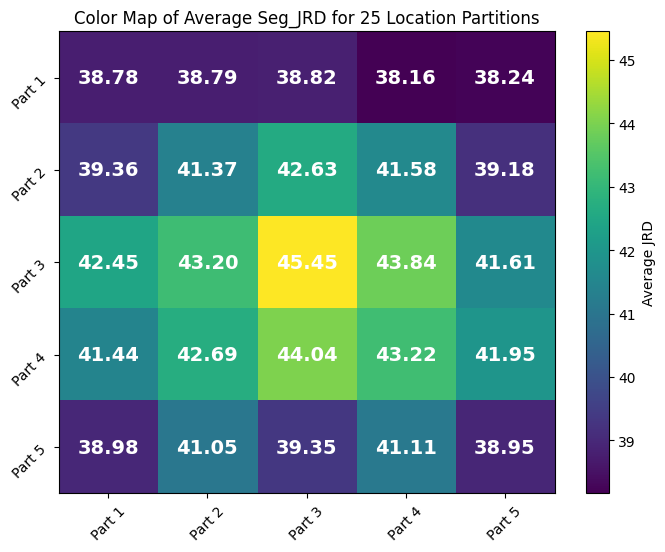}} 
        \subfloat[]{
        \includegraphics[width=0.16\textwidth]{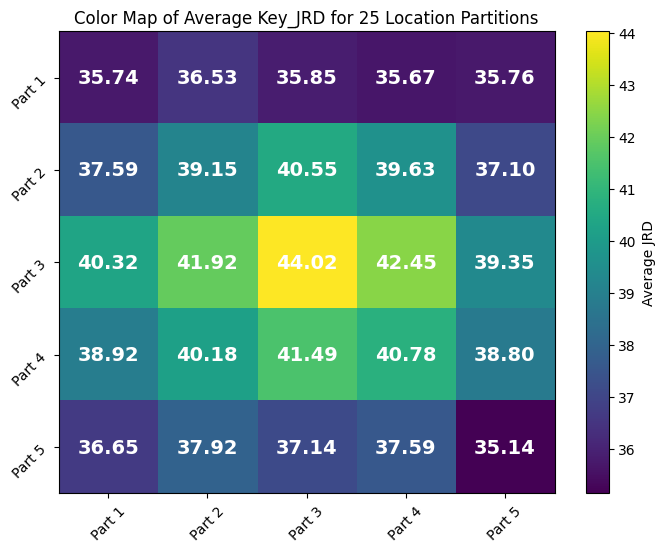}}
        \caption{Relationship between object location and JRD.(a) OD, (b) IS, (c) KPD}
	\label{MTJRD loc}
\end{figure}

Third, we adjust the threshold $T$ to validate its impact.
As $T$ increases from 0.65 to 0.85, the average JRD consistently decreases across all tasks, as illustrated in Fig. \ref{JRD_distribution}(b)-(d). For example, for OD, it decreases from 45.188 to 39.689.
Additionally, across all thresholds, OD consistently exhibits the highest JRD, followed by IS, with KPD remaining the lowest. 
These results confirm that the threshold $T$ effectively balances rate and distortion, with task-specific responses highlighting the inherent differences between machine vision tasks.

\subsubsection{Influencing Factors of JRD}
\label{section: JRD factors}
The size and position of an object are critical factors influencing the JRD. 
To investigate the relationship between object size and JRD, we sort all objects in the MT-JRD dataset by size in ascending order, divide them into ten equally populated intervals, and compute the average JRD within each interval.
As shown in Fig. \ref{MTJRD size}, a clear positive correlation between object size and JRD is observed across all three tasks. 
This trend occurs because larger objects provide more pixels, facilitating the extraction of richer semantic information. Consequently, distortion exerts a relatively smaller impact on the overall semantic integrity, thereby resulting in a higher JRD. 
The systematic increase in JRD for all tasks confirms that object size is a key factor influencing semantic rate distortion.

Fig. \ref{MTJRD loc} illustrates the relationship between the average JRD and object position. 
Quantitatively, for OD, the average JRD reaches 46.64 in the center region, while peripheral areas such as the top-left and bottom-right corners exhibit values of 39.72 and 39.26, respectively. A similar spatial trend is observed in both IS and KPD, where the central JRD exceeds the average JRD of the four corners by more than 6 and 8, respectively. 
This spatial disparity is inherently linked to the Effective Receptive Field (ERF) mechanism\cite{luo2016understanding}. Specifically, objects near the image center can aggregate richer global context through deep convolutional layers, yielding higher semantic redundancy. Consequently, these centrally located objects exhibit stronger robustness to distortion, thereby manifesting as a higher average JRD.

\begin{figure}[t]
    \centering
    \captionsetup{justification=justified}
        \subfloat[]{
        \includegraphics[width=0.45\textwidth]{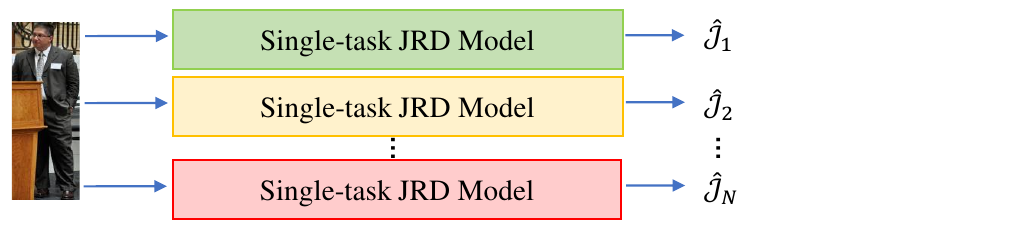}}\\
        \subfloat[]{
        \includegraphics[width=0.45\textwidth]{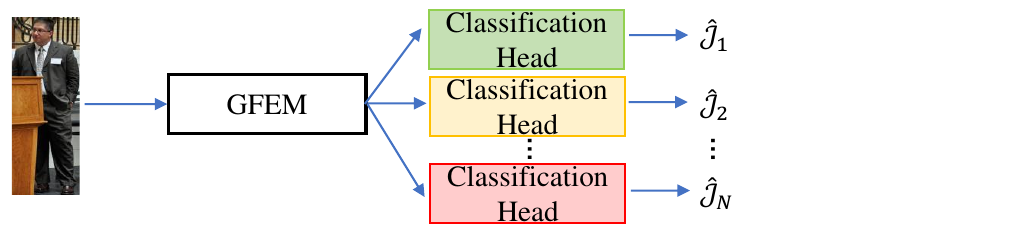}}\\
        \subfloat[]{
        \includegraphics[width=0.45\textwidth]{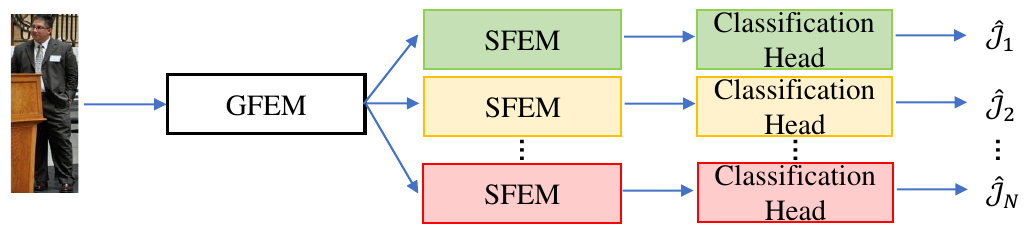}}
        \caption{Architectures of MT-JRD prediction models. (a) Independent single-task architecture, (b) Unified GFEM-based architecture, (c) Joint GFEM-SFEM architecture.}
	\label{arch}
\end{figure}

\section{Proposed AMT-JRD Prediction Model}
\label{section:framework}
\subsection{MT-JRD Prediction Frameworks}
\begin{figure*}[t]
	\centering
        \includegraphics[width=0.95\textwidth]{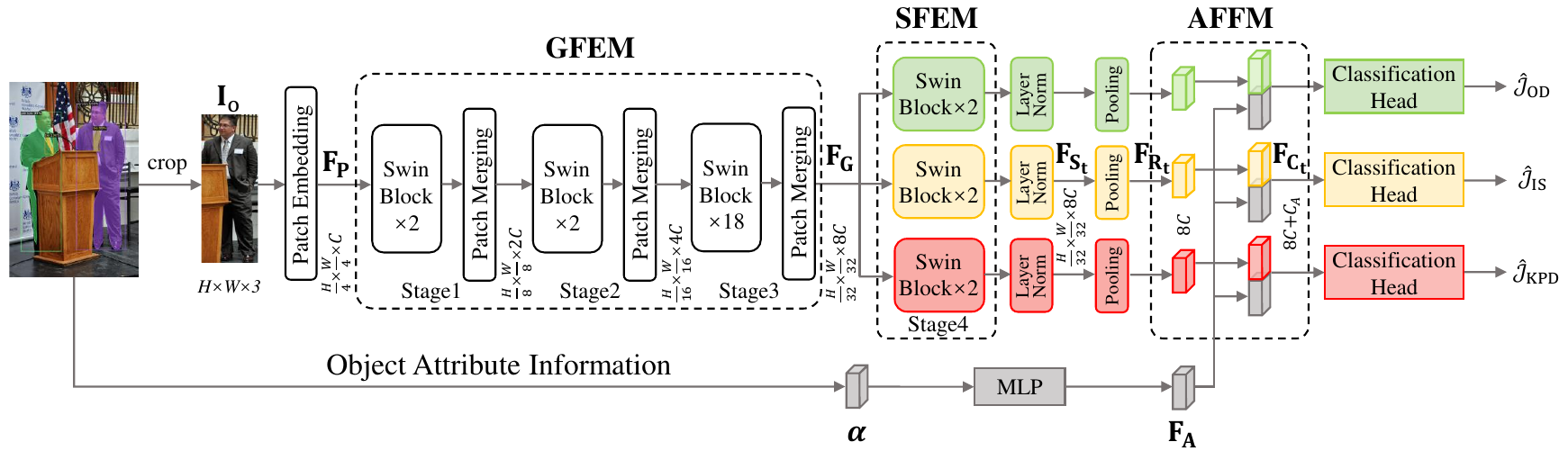}
        \captionsetup{justification=justified}
        \caption{The proposed AMT-JRD model, which consists of GFEM, SFEM, AFFM, and multi-task classification heads.}
        \label{JRD model}
\end{figure*}

The existing JRD models\cite{zhang2021just,zhang2023learning,liu2024dtjrd} were designed to predict JRD for a single task. As illustrated in Fig. \ref{arch} (a), if directly extending single-task JRD prediction to MT-JRD prediction, multiple independent prediction models are required to be trained, which are then performed and evaluated independently, i.e., one for each task. As a result, it leads to the highest number of parameters and computational complexity.
To address this issue, we propose a unified GFEM-based model, shown in Fig. \ref{arch} (b), which contains a GFEM module and multiple classification heads. This model predicts the JRD values for all three tasks in a single forward pass, resulting in the lowest parameter count and the shortest inference time. 
However, since it uses a shared GFEM to extract a single feature representation for all tasks, the same features are fed into the task-specific classification heads for the JRD prediction.
It reduces the capability of expressing task-specific characteristics and extracting task-dependent features.


To overcome this limitation in Fig. \ref{arch}(b), we propose a joint GFEM-SFEM model, as shown in Fig. \ref{arch}(c), which incorporates both the GFEM and SFEM to decouple features of different granularities. 
This hierarchical architecture follows the established paradigm of multi-task learning, where a shared backbone extracts common representations while specialized branches capture task-specific characteristics. It efficiently balances parameter efficiency with the capacity to model the unique demands of each task, which is essential for both accurate and effective MT-JRD prediction.

\subsection{Proposed Attribute-assisted MT-JRD}
Based on the analysis in Section \ref{section: JRD factors}, it reveals that the object size and location information are two key factors influencing JRD, which can be exploited in MT-JRD prediction.
Fig. \ref{JRD model} shows the proposed Attribute-assisted MT-JRD prediction model, named AMT-JRD, which consists of a shared GFEM, task-specific SFEM branches, an AFFM, and task-specific heads for OD, IS, and KPD tasks. The AMT-JRD is implemented by a Swin-Transformer \cite{liu2021swin} backbone, where stages 1 to 3 serve as the shared GFEM and stage 4 is duplicated as three task-specific SFEMs. The AFFM combines attributes as complementary features for JRD prediction.

For an original image $\mathbf{I}_{\mathrm{o}}\in\mathbb{R}^{H \times W \times 3}$, it is first transformed into patch tokens $\mathbf{F_P}\in \mathbb{R}^{\frac{H}{4} \times \frac{W}{4} \times C}$ through the embedding layer. Then, GFEM captures generalized features $\mathbf{F_G}\in \mathbb{R}^{\frac{H}{32} \times \frac{W}{32} \times 8C}$ for all tasks. After that, $\mathbf{F_G}$ is input to three SFEMs, followed by a layernorm at the tail. The specialized features $\mathbf{F_{S_t}}\in \mathbb{R}^{\frac{H}{32} \times \frac{W}{32} \times 8C}$ are aggregated by an average pooling to generate refined image features $\mathbf{F_{R_t}}\in \mathbb{R}^{8C}$. The process is presented as 

\begin{equation}
\left\{
\begin{aligned}
\mathbf{F_G} &= \Phi_{\mathrm{GFEM}}(\Phi_{\mathrm{Emb}}(\mathbf{I}_{\mathrm{o}})) \\
\mathbf{F_{S_t}} &= \Phi_{\mathrm{LN}_t}(\Phi_{\mathrm{SFEM}_t}(\mathbf{F_G})) \\
\mathbf{F_{R_t}} &= \Phi_{\mathrm{AP}}(\mathbf{F_{S_t}})
\end{aligned}
\right.
\label{model_equations}
\end{equation}
where $\Phi_{\mathrm{Emb}}$ represents the embedding layer, $\Phi_{\mathrm{LN}_t}$ and $\Phi_{\mathrm{SFEM}_t}$ denote layernorm and SFEM module for task $t$,
$t\in\{\text{OD},\text{IS},\text{KPD}\}$, and $\Phi_{\mathrm{AP}}$ is the average pooling operation.

The attributes, such as object size and location, are useful, but cannot be directly captured by GFEM and SFEM. Therefore, we propose the AFFM, where attributes serve as complementary features for image representations.
For an object of width $w$ and height $h$, its size is normalized as \( s=\frac{w \times h}{224^2} \) to prevent excessively large values from causing instability during training. The object location is represented by its center coordinates \( (x_0, y_0) \). As a result, we obtain an attribute triplet $\bm{\alpha}=(s, x_0, y_0)$. Then, $\bm{\alpha}$ is projected by an MLP into a higher-dimensional attribute feature $\mathbf{F_A} \in \mathbb{R}^{C_A}$, which increases representational capacity.
$\mathbf{F_A}$ is then concatenated with the $\mathbf{F_{R_t}}$ to form complementary features $\mathbf{F_{C_t}}\in \mathbb{R}^{8C+C_A}$, enabling the attribute information to be embedded into the extracted visual representations.
The AFFM module introduces object prior information, compensating for the image features. This information enhances the AMT-JRD model in perceptual modeling of machine vision.

Finally, MT-JRDs $\hat{J}_t$ are predicted as

\begin{equation}
\left\{
\begin{aligned}
\mathbf{F_A} &= \Phi_{\mathrm{MLP}}(\boldsymbol{\alpha}) \\
\mathbf{F_{C_t}} &= \mathbb{C}(\mathbf{F_{R_t}}, \mathbf{F_A}) \\
\hat{J}_t &= \Phi_{\mathrm{H}_t}(\mathbf{F_{C_t}})
\label{head}
\end{aligned}
\right.
\end{equation}
where $\mathbb{C}(\cdot)$ denotes feature concatenation operation, $\Phi_{\mathrm{H}_t}$ represents the classification head for task $t$, and $\hat{J}_t$ indicates the predicted JRD for task $t$.

\begin{figure}[t]
	\centering
        \subfloat[]{
        \includegraphics[width=0.42\textwidth]{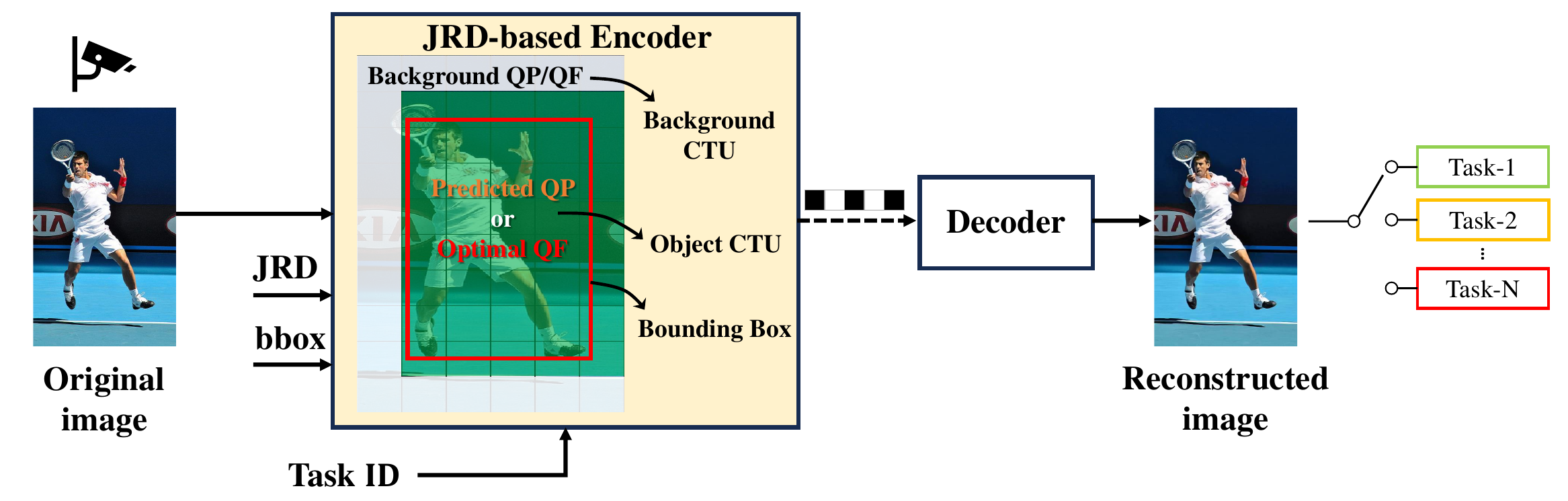}}\\
        \subfloat[]{
        \includegraphics[width=0.4\textwidth]{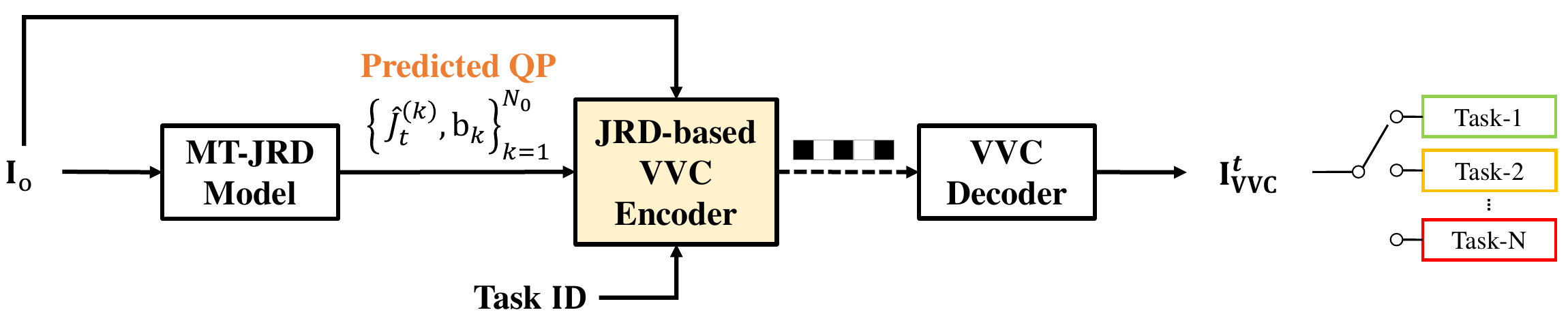}}\\
        \subfloat[]{
        \includegraphics[width=0.48\textwidth]{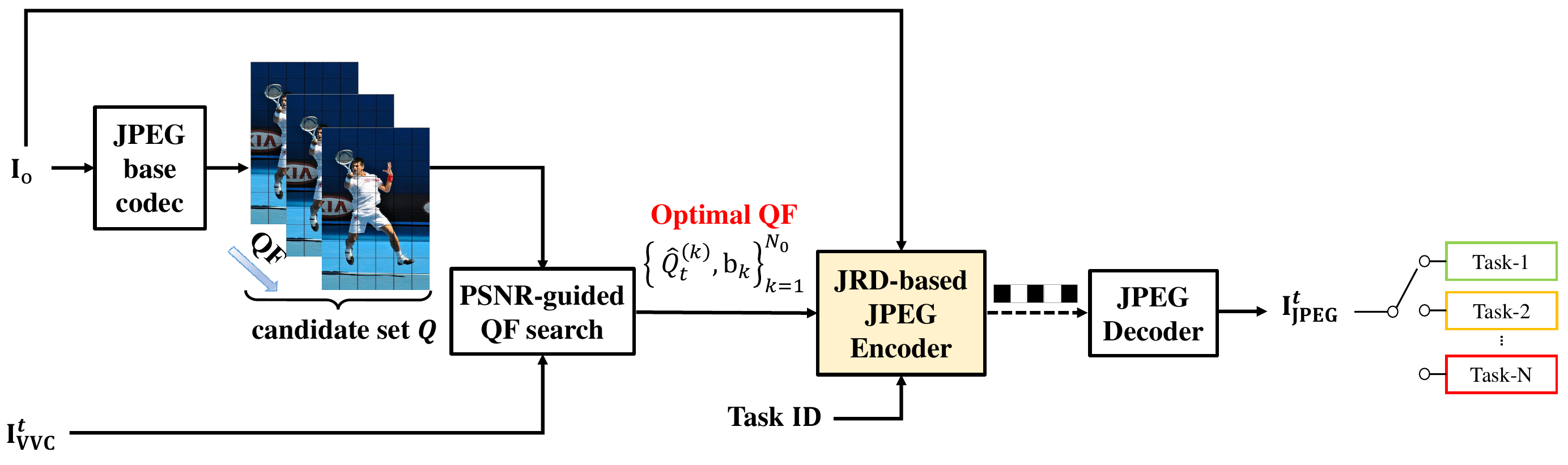}}
        \captionsetup{justification=justified}
        \caption{The processing workflow for JRD-based VCM optimization. (a) JRD-based VCM\cite{liu2024dtjrd}, (b) JRD-based VVC\cite{liu2024dtjrd}, (c) JRD-based JPEG.}
        \label{VCM}
\end{figure}

\section{MT-JRD-based VCM}
\label{section:vcm}

To fully exploit the JRD property in visual communication, we propose an MT-JRD-based VCM to reduce the bit rate while maintaining multi-task machine vision accuracies. 
As illustrated in Fig. \ref{VCM}(a), the JRD-based VCM is task-switchable and compatible with different base codecs, such as JPEG and VVC.
Its inputs include the original image $\mathbf{I}_{\mathrm{o}}$, the JRD values of individual objects, and their bounding boxes $\{\mathbf{b}_k\}_{k=1}^{N_\mathrm{o}}$, where $N_\mathrm{o}$ is the total number of objects in $\mathbf{I}_{\mathrm{o}}$. 
The JRD values are represented by coding control parameters such as the QP in VVC or the Quantization Factor (QF) in JPEG. The core of the proposed JRD-based VCM method is to utilize the object-wise JRD values to realize spatially adaptive quantization, i.e., differentiated compression ratios of foreground objects and background regions.

As shown in Fig. \ref{VCM}(b) and (c), JRD-based VCM is instantiated twice, corresponding to the VVC and JPEG coding schemes, respectively.
First, we follow the JRD-based VCM proposed in \cite{liu2024dtjrd}, i.e., the JRD-based VVC, to realize spatially adaptive quantization. The VVC-reconstructed image optimized for task $t$, i.e., $\mathbf{I}_{\text{VVC}}^t$, serves as the quality reference for the subsequent optimization process.

Second, as both the dataset and model training of AMT-JRD are developed based on VVC, the proposed JRD-based VCM can be extended to other codecs, such as JPEG, to support a more generalized VCM. 
We use JPEG base codec to generate a candidate set $Q$ consisting of a series of QF values.
Given $\mathbf{I}_{\mathrm{o}}$, $\mathbf{I}_{\text{VVC}}^t$, and $\{\mathbf{b}_k\}_{k=1}^{N_\mathrm{o}}$, the objective is to determine the optimal QF $\hat{Q}_{t}^{(k)}$ for each bounding box $\mathbf{b}_k$ in task $t$ such that the JPEG reconstruction quality matches the VVC reconstruction PSNR quality as closely as possible, which is defined as


\begin{equation}
\left\{
\begin{aligned}
D(QF) &=
\left|
\text{PSNR}\!\left(\mathbf{I}_\mathrm{o}^{b_k}, \mathbf{I}_{\text{JPEG}(QF)}^{b_k}\right)
-
\text{PSNR}\!\left(\mathbf{I}_\mathrm{o}^{b_k}, \mathbf{I}_{\text{VVC}}^{t,b_k}\right)
\right|, \\[4pt]
\hat{Q}_{t}^{(k)} &= \arg\min_{QF \in Q} D(QF)
\end{aligned}
\right.
\label{QF_search}
\end{equation}
where $\mathbf{I}_\mathrm{o}^{b_k}$, $\mathbf{I}_{\text{JPEG}(QF)}^{b_k}$, and $\mathbf{I}_{\text{VVC}}^{t,b_k}$ denote regions within the bounding box of the original image, the JPEG-reconstructed image with uniform $QF$, and the VVC-reconstructed image optimized for task $t$, respectively.
The time complexity can be reduced to $O(log_{2}{n})$ by applying binary search in the candidate set $Q$. After deriving the optimal QF $\hat{Q}_{t}^{(k)}$ for all bounding boxes, a block-level QF map is constructed and integrated into the JPEG encoder, thereby enabling JRD-based JPEG compression and obtaining the final JPEG-reconstructed image optimized for task $t$, i.e., $\mathbf{I}_{\text{JPEG}}^t$.

\section{Experimental Results and Analysis}
\label{section:experiment}

\begin{table*}[htbp]
\centering
\caption{Average accuracy and standard deviation comparison between the proposed AMT-JRD and the state-of-the-art JRD models.}
\renewcommand{\arraystretch}{1.2}
\setlength{\tabcolsep}{6pt}
\begin{tabular}{c|ccc|ccc|ccc|ccc}
\toprule
\multirow{2}{*}{\textbf{Model}} &
\multicolumn{3}{c|}{\textbf{OD}} &
\multicolumn{3}{c|}{\textbf{IS}} &
\multicolumn{3}{c|}{\textbf{KPD}} &
\multicolumn{3}{c}{\textbf{Avg.}} \\
\cmidrule(lr){2-4} \cmidrule(lr){5-7} \cmidrule(lr){8-10} \cmidrule(lr){11-13}
 & $E_A$ & $E_{[27,51]}$ & $\sigma_e$ 
 & $E_A$ & $E_{[27,51]}$ & $\sigma_e$ 
 & $E_A$ & $E_{[27,51]}$ & $\sigma_e$ 
 & $E_A$ & $E_{[27,51]}$ & $\sigma_e$ \\
\midrule
EL-JRD (IJCV 21’)\cite{zhang2021just}  & 5.933 & 5.721 & 7.949 & 5.613 & 5.122 & 7.495 & 5.778 & 5.215 & 7.651 & 5.775 & 5.353 & 7.698 \\
BC-JRD (TMM 23’)\cite{zhang2023learning}  & 4.555 & 4.283 & 6.572 & 4.326 & 3.875 & 5.954 & 4.276 & 3.866 & 6.297 & 4.386 & 4.008 & 6.274 \\
DT-JRD (TMM 25’)\cite{liu2024dtjrd} & 3.929 & 3.442 & 5.564 & 4.001 & 3.428 & 5.648 & 4.225 & 3.632 & 5.867 & 4.052 & 3.501 & 5.693 \\
MT-JND (ICIP 23’)\cite{10222099} & 4.586 & 4.022 & 6.102 & 4.757 & 4.113 & 6.302 & 4.637 & 3.868 & 6.387 & 4.660 & 4.001 & 6.264 \\
MJ-REC (TCSVT 24’)\cite{10500870} & 4.709 & 4.182 & 6.276 & 4.863 & 4.235 & 6.440 & 5.080 & 4.282 & 6.823 & 4.884 & 4.233 & 6.513 \\
AMT-JRD (Ours) & \textbf{3.796} & \textbf{3.395} & \textbf{5.415} 
& \textbf{3.818} & \textbf{3.337} & \textbf{5.365} 
& \textbf{3.728} & \textbf{3.286} & \textbf{5.217} 
& \textbf{3.781} & \textbf{3.339} & \textbf{5.332} \\
\bottomrule
\end{tabular}
\label{tab:multi-task-eval}
\end{table*}

\begin{table*}[t]
    \centering
    \caption{Multi-task accuracy comparison in terms of differences and $R^2$ of PSNR [dB], SSIM and LPIPS.}
    \label{quality performance}
    \renewcommand{\arraystretch}{1.2} 
    \setlength{\tabcolsep}{3pt} 
    \begin{tabular}{ccccccc}
        \bottomrule
        Model & EL-JRD\cite{zhang2021just} & BC-JRD\cite{zhang2023learning} & DT-JRD\cite{liu2024dtjrd} & MT-JND\cite{10222099} & MJ-REC\cite{10500870} & AMT-JRD\\
        \hline
        $\lvert\Delta$PSNR$\rvert$ $\downarrow$ & 0.617\,/\,0.686\,/\,0.745 & 0.502\,/\,0.535\,/\,0.482 & 0.485\,/\,0.502\,/\,0.545 & 0.559\,/\,0.611\,/\,0.612 & 0.579\,/\,0.606\,/\,0.665 & \textbf{0.441}\,/\,\textbf{0.468}\,/\,\textbf{0.450} \\
        $\lvert\Delta$SSIM$\rvert(\times 10^{-3})$ $\downarrow$ & 9.693\,/\,11.052\,/\,11.584 & 7.921\,/\,8.618\,/\,8.237 & 7.655\,/\,8.143\,/\,8.876 & 8.983\,/\,10.237\,/\,9.629 & 9.207\,/\,9.966\,/\,10.761 & \textbf{6.916}\,/\,\textbf{7.770}\,/\,\textbf{7.476} \\
        $\lvert\Delta$LPIPS$\rvert(\times 10^{-3})$ $\downarrow$ & 5.297\,/\,5.851\,/\,5.761 & 4.413\,/\,4.532\,/\,4.402 & 4.382\,/\,4.339\,/\,4.440 & 5.184\,/\,5.484\,/\,4.673 & 5.278\,/\,5.203\,/\,5.310 & \textbf{4.176}\,/\,\textbf{4.151}\,/\,\textbf{4.099} \\
        $R^2$ of PSNR $\uparrow$ & 0.921\,/\,0.914\,/\,0.870 & 0.948\,/\,0.942\,/\,0.955 & 0.952\,/\,0.951\,/\,0.932 & 0.939\,/\,0.930\,/\,0.917 & 0.934\,/\,0.936\,/\,0.901 & \textbf{0.961}\,/\,\textbf{0.954}\,/\,\textbf{0.961} \\
        $R^2$ of SSIM $\uparrow$ & 0.953\,/\,0.935\,/\,0.918 & 0.971\,/\,0.957\,/\,0.960 & 0.971\,/\,0.963\,/\,0.952 & 0.959\,/\,0.945\,/\,0.951 & 0.959\,/\,0.963\,/\,0.935 & \textbf{0.978}\,/\,\textbf{0.967}\,/\,\textbf{0.967} \\
        $R^2$ of LPIPS $\uparrow$ & 0.898\,/\,0.868\,/\,0.876 & 0.932\,/\,0.919\,/\,0.925 & 0.931\,/\,0.927\,/\,0.923 & 0.904\,/\,0.892\,/\,0.932 & 0.900\,/\,0.900\,/\,0.901 & \textbf{0.945}\,/\,\textbf{0.936}\,/\,\textbf{0.936}\\
        \toprule
    \end{tabular}
\end{table*}

\subsection{Experimental Settings}
\subsubsection{Dataset and Model Configurations}
Images in the JRD dataset were randomly divided into training, validation, and test sets in an 8:1:1 ratio. 
Images were resized to 224$\times$224 resolution and horizontally flipped with a probability of 0.5 during training.
The AMT-JRD model adopted the pre-trained Swin Transformer-small backbone, with embedding dimension $C$ and attribute feature dimension $C_A$ set to 96 and 256, respectively.
All the models were trained on a workstation equipped with NVIDIA RTX 3090 GPUs. 
The batch size was set as 32, and the learning rate was initially set to 0.01 and adjusted using the cosine decay strategy. 
The GDSL-based loss proposed in \cite{liu2024dtjrd} was used, and the standard deviation of the Gaussian distribution was set as 3.

\subsubsection{Benchmark JRD Configurations}
We compared AMT-JRD with other state-of-the-art JRD prediction models, including ``EL-JRD''\cite{zhang2021just}, ``BC-JRD''\cite{zhang2023learning}, ``DT-JRD''\cite{liu2024dtjrd}, ``MT-JND''\cite{10222099} and ``MJ-REC''\cite{10500870}. MT-JND and MJ-REC are methods designed for modeling the JND for HVS. EL-JRD and BC-JRD are designed for modeling JRD in machine vision. However, as they are based on multiple binary classification frameworks, they are limited to single-task prediction. DT-JRD was specifically proposed for JRD and achieves excellent performance. It was originally designed for a single OD task, but can be extended to multi-task prediction in the form of GFEM design.

\subsubsection{Multi-task VCM Configurations}
For VVC, all-intra configuration was used as the base encoder and $QP \in \{30, 31, 32, 33, 34\}$ for three tasks. Regarding JRD-based VVC, the predicted JRDs $\hat{J}_t^{(k)}$ from six JRD models and the ground truth JRD ${J}_t^{(k)}$ were set as QPs for the foreground objects. To balance the bit rate and multi-task machine vision performance, the foreground objects were encoded with JRDs plus a QP offset, i.e., $\hat{J}_t^{(k)}+\Delta QP$ or ${J}_t^{(k)}+\Delta QP$, where $\Delta QP$ $\in$ \{-4, -3, -2, -1, 0\}. The background region was encoded with a larger background QP, e.g., 51\cite{zhang2023learning}.

For JPEG, the base encoder was configured with uniform compression ratio, i.e., $QF$ $\in$ \{42, 44, 46, 48, 50\} for OD and IS tasks, and $QF$ $\in$ \{46, 48, 50, 52, 54\} for KPD task. Regarding JRD-based JPEG, the optimal QFs $\hat{Q}_{t}^{(k)}$ and ground truth QF ${Q}_{t}^{(k)}$ were first derived by PSNR-guided QF search shown in Fig.\ref{VCM}(c) and then applied to the foreground objects. Similarly, the foreground objects were encoded with QFs plus a QF offset, i.e., $\hat{Q}_{t}^{(k)}+\Delta QF$ or ${Q}_{t}^{(k)}+\Delta QF$, where $\Delta QF$ $\in$ \{0, 1, 2, 3, 4\}. The background region was encoded with a smaller background QF, e.g., 30, in this work.


\subsubsection{Evaluation Metrics}
{(1) JRD prediction accuracy:}
First, the global and local Mean Absolute Error (MAE) between the predicted JRD and the ground truth JRD are calculated\cite{zhang2023learning}.
The MAE over all samples is denoted as $E_A$, and the MAE over samples whose ground-truth JRD falls within [27, 51] is denoted as $E_{[27, 51]}$.
Second, to reflect the model's consistency
and stability, we compute the standard deviation of error $\sigma_e$\cite{liu2024dtjrd}.
Third, quality differences and their $R^2$ were also evaluated.
$\lvert\Delta$PSNR$\rvert$ calculates the MAE of the PSNR values between images compressed with the ground truth JRD and those compressed with the predicted JRD. $R^2$ of PSNR is defined as the PSNR correlation between the predicted and ground truth JRDs. Similar procedure is applied for both SSIM\cite{SSIM} and LPIPS\cite{LPIPS}.
{(2) Coding performance:} Coding gain is evaluated with rate-accuracy curves, where accuracy refers to the mAP calculated using the respective similarity measures across three machine vision tasks. Also, we present visualization results and computational complexity analysis.




\begin{table*}[t]
\centering
\caption{Accuracy comparison of MT-JRD components in three tasks.}
\renewcommand{\arraystretch}{1.2}
\setlength{\tabcolsep}{5pt}
\begin{tabular}{cc|ccc|ccc|ccc|ccc}
\toprule
\multicolumn{2}{c|}{\multirow{2}{*}{\textbf{Architecture}}} &  
\multicolumn{3}{c|}{\textbf{OD}} & 
\multicolumn{3}{c|}{\textbf{IS}} & 
\multicolumn{3}{c|}{\textbf{KPD}} & 
\multicolumn{3}{c}{\textbf{Avg.}} \\
\cmidrule(lr){3-5} \cmidrule(lr){6-8} \cmidrule(lr){9-11} \cmidrule(lr){12-14}
& & $E_A$ & $E_{[27,51]}$ & $\sigma_e$ 
  & $E_A$ & $E_{[27,51]}$ & $\sigma_e$ 
  & $E_A$ & $E_{[27,51]}$ & $\sigma_e$ 
  & $E_A$ & $E_{[27,51]}$ & $\sigma_e$ \\
\midrule
Single-task           & --    & 3.947 & 3.560 & 5.503 & 4.016 & 3.586 & 5.515 & 3.851 & 3.438 & 5.353 & 3.938 & 3.528 & 5.457 \\
\midrule
\multirow{3}{*}{Multi-task}
& GFEM                 & 3.957 & 3.564 & 5.530 & 3.994 & 3.525 & 5.462 & 4.017 & 3.561 & 5.512 & 3.989 & 3.550 & 5.501 \\
& GFEM+SFEM            & 3.855 & 3.459 & 5.409 & 3.900 & 3.445 & 5.431 & 3.816 & 3.406 & 5.338 & 3.857 & 3.437 & 5.393 \\
& GFEM+SFEM+AFFM       & 3.796 & 3.395 & 5.415
                       & 3.818 & 3.337 & 5.365 
                       & 3.728 & 3.286 & 5.217
                       & \textbf{3.781} & \textbf{3.339} & \textbf{5.332} \\
\bottomrule
\end{tabular}
\label{tab:ablation-architecture}
\end{table*}
\subsection{MT-JRD Prediction Accuracy Evaluation}

To evaluate the effectiveness of the proposed AMT-JRD model, we conduct a prediction accuracy comparison between the AMT-JRD and the state-of-the-art methods, as shown in Table \ref{tab:multi-task-eval}. 
MT-JND and MJ-REC exhibit relatively large $E_A$ values of 4.660 and 4.884, respectively.
Since these two methods were originally designed to model HVS characteristics, they are suboptimal when transferred to machine vision tasks.
EL-JRD and BC-JRD reach $E_A$ values of 5.775 and 4.386, respectively.
Their need for multiple binary decisions introduces prediction inconsistency and error accumulation, resulting in inferior accuracy.
DT-JRD yields superior results over previous methods with an $E_A$ of 4.502 and $\sigma_e$ of 5.693.
The proposed GDSL extends the number of JRD labels and contributes to precise single-task JRD prediction.
Nevertheless, DT-JRD lacks joint modeling across tasks, thus limiting its generalization to multi-task settings.
In comparison, the proposed AMT-JRD achieves $E_A$ of 3.781, $E_{[27,51]}$ of 3.339 and $\sigma_e$ of 5.332, outperforming those of the DT-JRD by 6.7\%, 4.6\% and 6.3\% respectively. These significant and consistent improvements in all three tasks stem from our multi-branch architecture incorporating the GFEM, SFEM, and AFFM, which together enable accurate and robust MT-JRD modeling across diverse machine tasks.

In addition, we evaluate the 
accuracy and correlation
of the MT-JRD prediction using three metrics commonly used in image quality assessment, i.e., PSNR, SSIM\cite{SSIM}, and LPIPS\cite{LPIPS}. 
Table \ref{quality performance} shows that AMT-JRD achieves $\lvert\Delta$PSNR$\rvert$ of 0.441, 0.468, and 0.450 for OD, IS, and KPD tasks, respectively, outperforming all five other JRD prediction models.
The $R^2$ values for PSNR reach as high as 0.961, 0.954, and 0.961, underscoring the superior precision and concentration of the model's predictions.
Similarly, for both SSIM and LPIPS, the AMT-JRD demonstrates consistent superiority and achieves the best across all three tasks.
The fact that AMT-JRD achieves both the lowest MAE and the highest $R^2$ jointly demonstrates that its compressed image quality more closely approximates that of the ground truth JRD. The multi-task design of AMT-JRD enhances the protection of generalized-to-specialized perceptual features, which enable more accurate and robust modeling of JRD thresholds.


\subsection{Ablation Studies on AMT-JRD}

\begin{figure*}[t]
    \centering
        \subfloat[]{
        \includegraphics[width=0.32\textwidth]{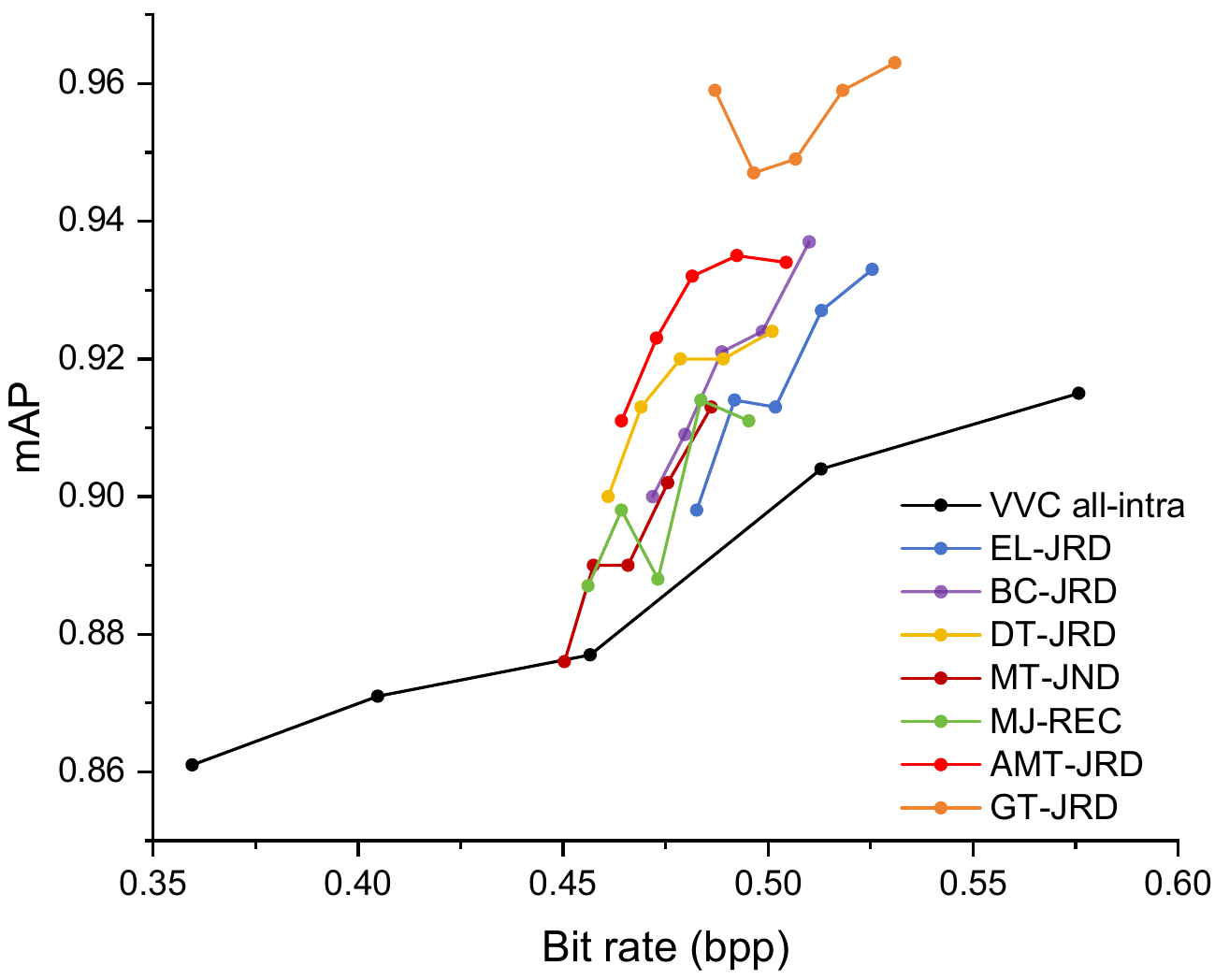}
        \label{VVC_det}}
        \subfloat[]{
        \includegraphics[width=0.32\textwidth]{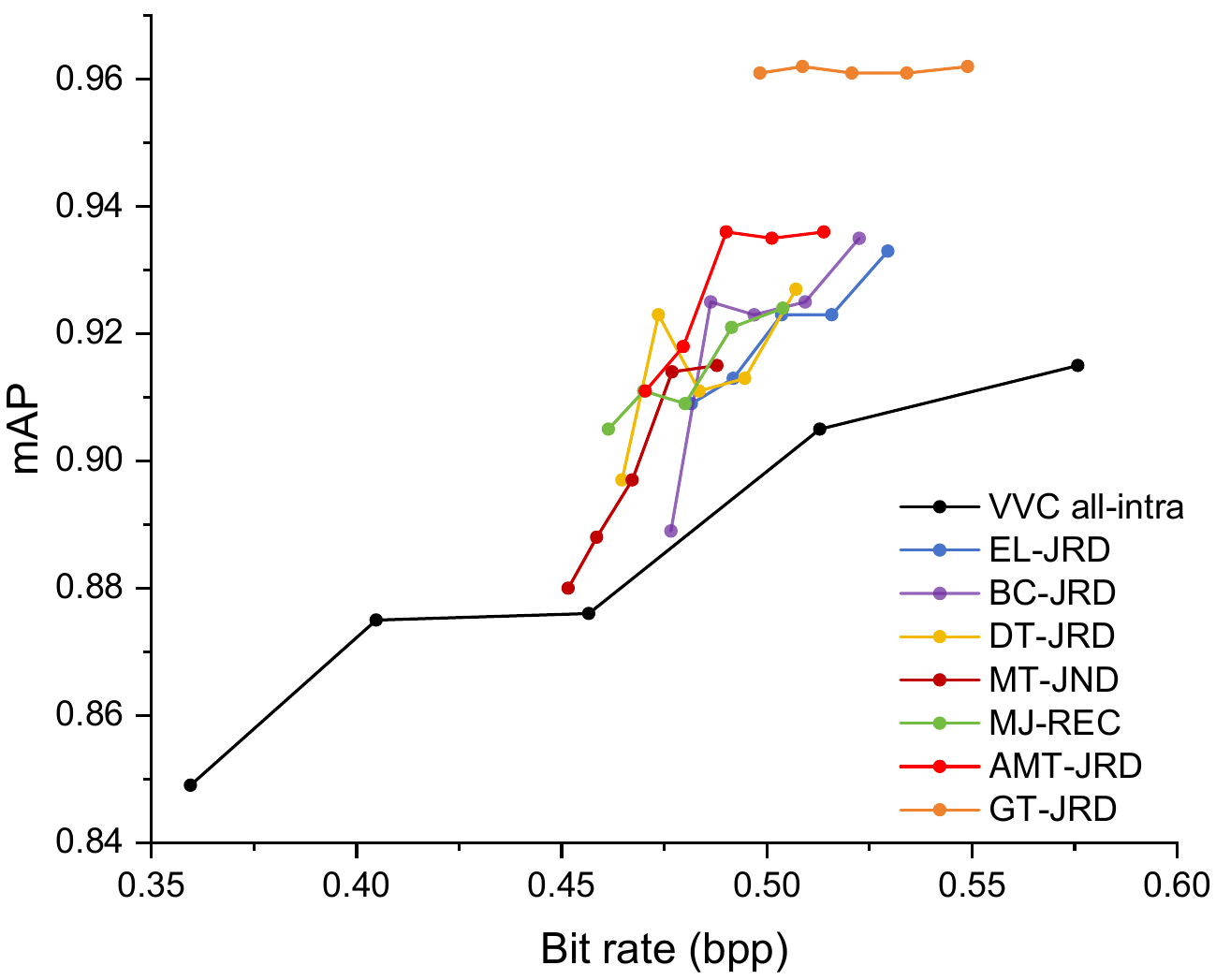}
        \label{VVC_seg}}
        \subfloat[]{
        \includegraphics[width=0.32\textwidth]{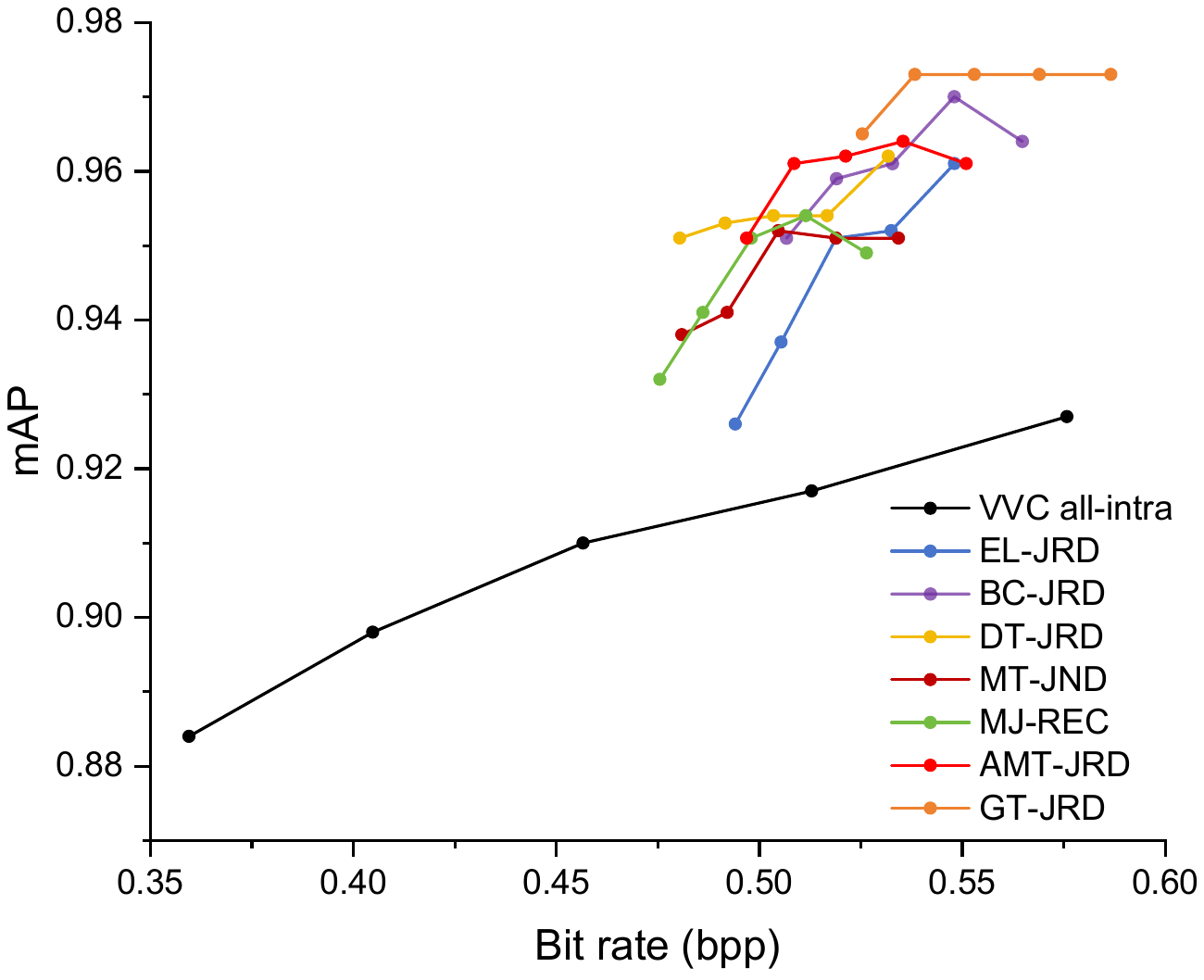}
        \label{VVC_key}}
    \captionsetup{justification=justified}
    \caption{
    Coding gain comparison among JRD models in VVC.
    (a) OD, (b) IS, (c) KPD.}
    \label{VVC}
\end{figure*}

\begin{figure*}[t]
    \centering
        \subfloat[]{
        \includegraphics[width=0.32\textwidth]{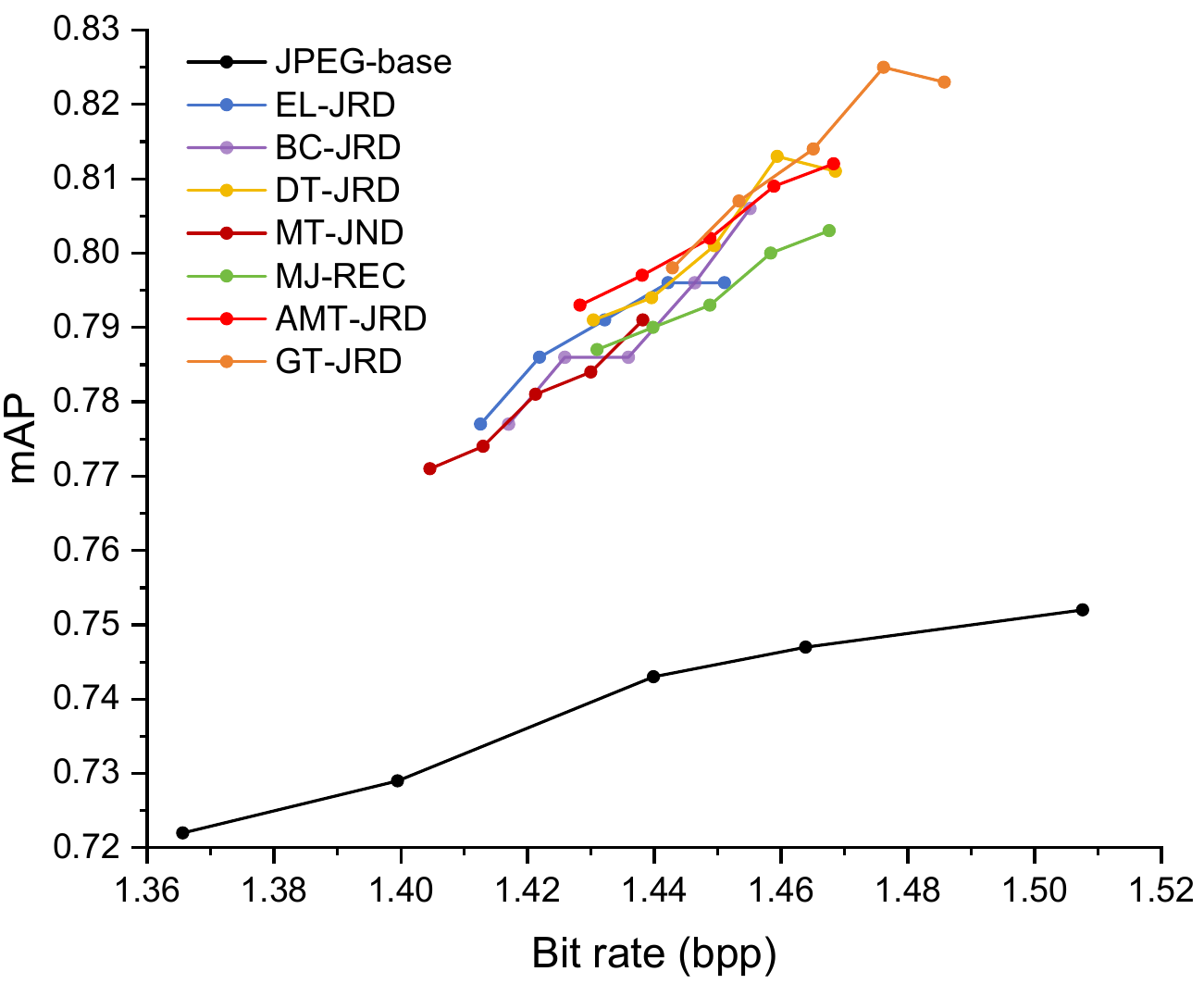}
        \label{VVC_det}}
        \subfloat[]{
        \includegraphics[width=0.32\textwidth]{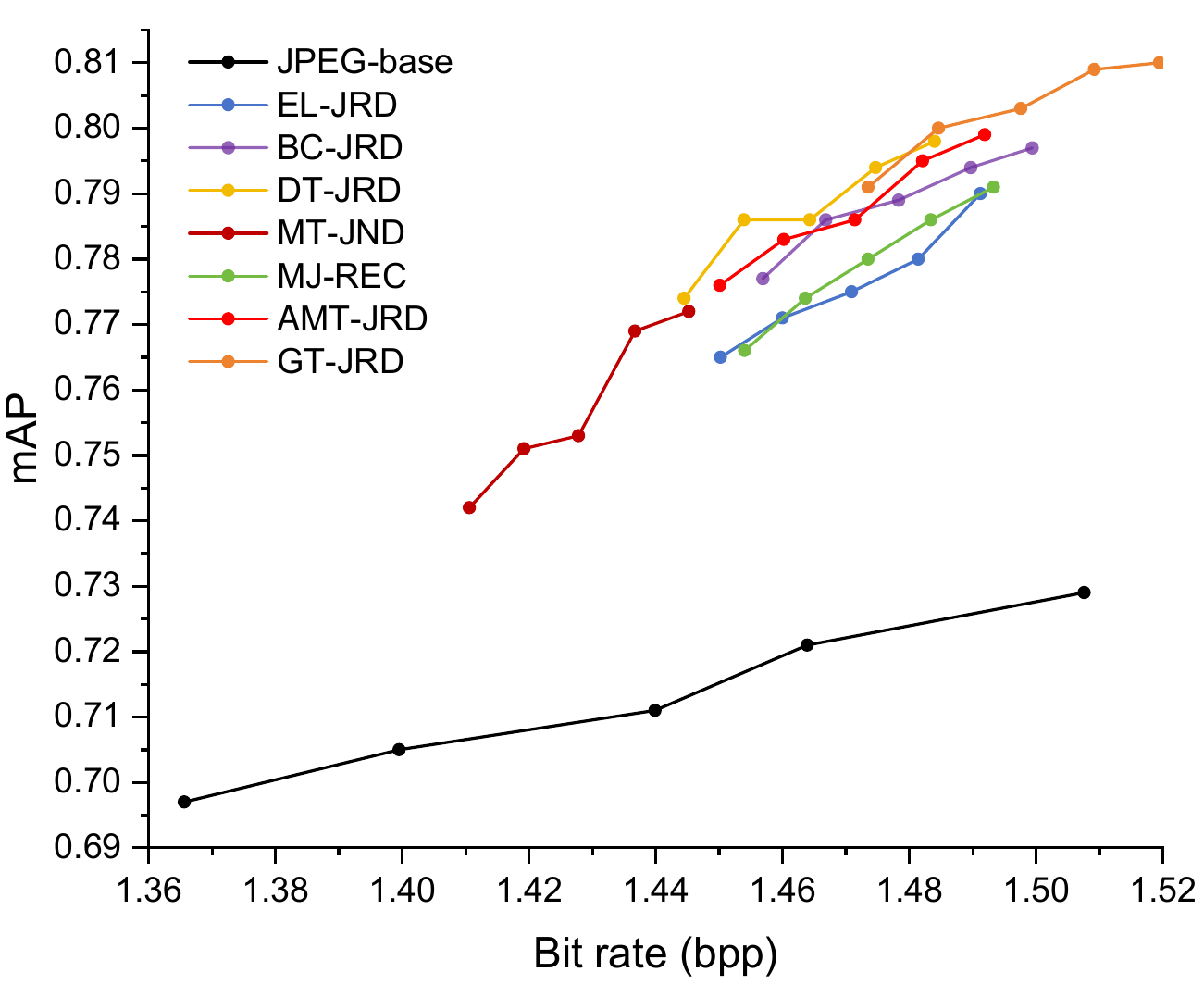}
        \label{VVC_seg}}
        \subfloat[]{
        \includegraphics[width=0.32\textwidth]{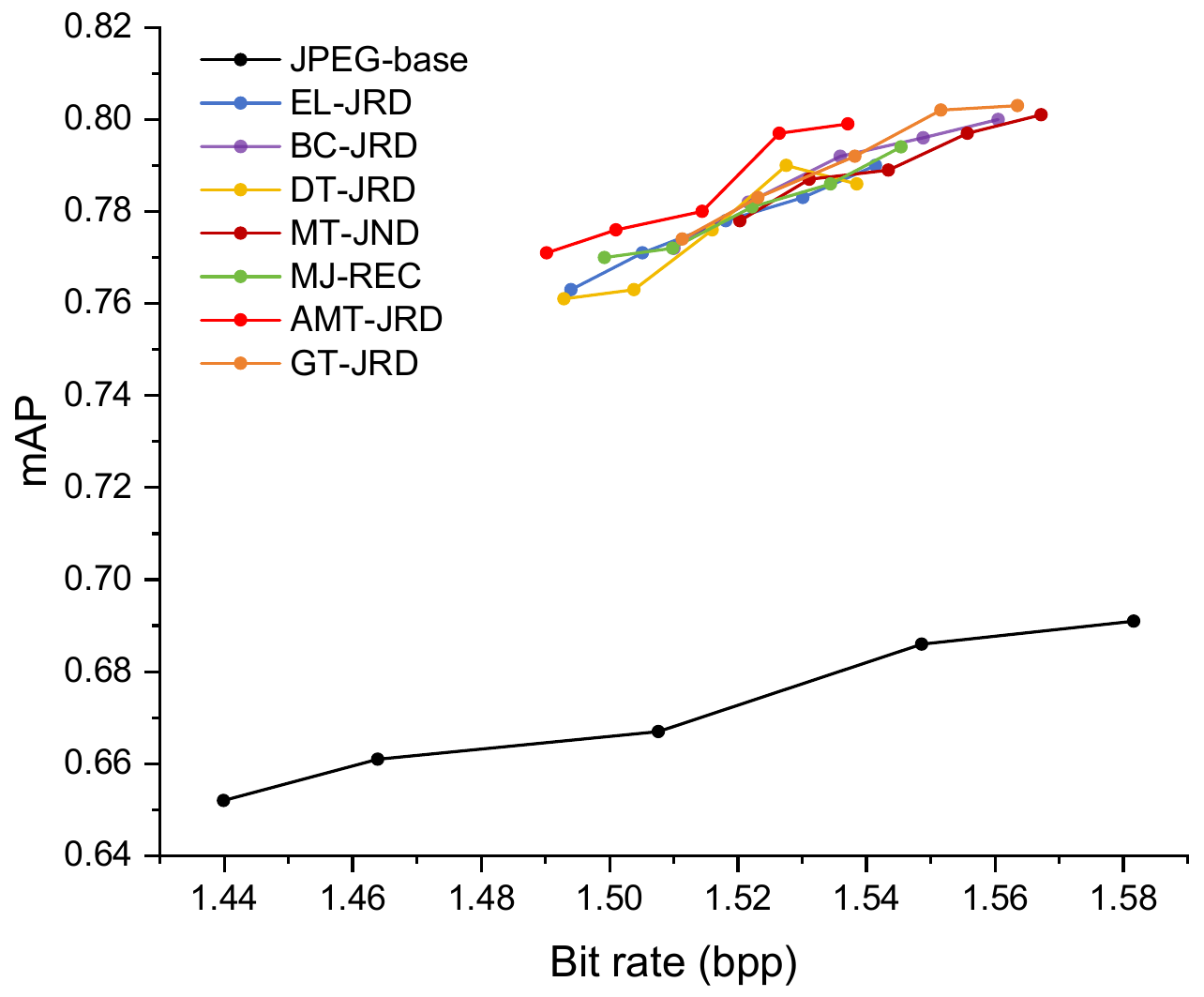}
        \label{VVC_key}}
    \captionsetup{justification=justified}
    \caption{Coding gain comparison among JRD models in JPEG. (a) OD, (b) IS, (c) KPD. }
    \label{JPEG}
\end{figure*}

To investigate the effectiveness of each component in the proposed AMT-JRD, an ablation study was performed by progressively incorporating the GFEM, SFEM, and AFFM modules. Table \ref{tab:ablation-architecture} shows the accuracy comparison among different combinations of the GFEM, SFEM, and AFFM modules in MT-JRD.
For the single-task model, the architecture shown in Fig. \ref{arch}(a) performs as the baseline.
For the multi-task model, when only GFEM is used, it reaches the highest $E_A$ of 3.989.
This result underscores the limitation of using a shared feature extractor without explicitly modeling task-specific variations. 
Although global context aggregation provides a unified feature backbone, it lacks the capacity to generate sufficiently diverse representations for heterogeneous tasks.
Upon incorporating the SFEM, the model exhibits consistent improvements, with $E_A$, $E_{[27,51]}$, and $\sigma_e$ each reduced by more than 0.1 compared to the GFEM-only model.
This demonstrates that explicitly learning task-aware representations is crucial for distinguishing perceptual boundaries in OD, IS, and KPD.
Further accuracy gains are achieved when AFFM is added. Compared to the single-task baseline, the full AMT-JRD model achieves approximately 4.0\%, 5.4\% and 2.3\% reduction in $E_A$, $E_{[27,51]}$, and $\sigma_e$.
By injecting auxiliary information such as object size and spatial position, AFFM enhances contextual understanding of the model and leads to the most accurate and stable predictions. 
These results confirm that GFEM is a strong but generic foundation, while SFEM and AFFM progressively improve task specificity and attribute awareness, jointly enabling precise and robust MT-JRD prediction.

\subsection{Coding Efficiency Evaluation on Multiple Machine Tasks}

In this study, JRD-based VCM is implemented using both VVC and JPEG as base codecs. Figs. \ref{VVC} and \ref{JPEG} show the rate-accuracy of the JRD-based VCM on three representative tasks. The horizontal axis in each subplot represents the bit rate in bits per pixel (bpp), and the vertical axis indicates the mAP. 
Given that the proposed MT-JRD dataset is constructed based on the semantic content of person, mAP is effectively equivalent to the average precision for the person category.
For the three tasks, mAP is computed using the overlap of the bounding box for OD, the mask intersection for IS, and the similarity of the keypoint vector for KPD, respectively.

\begin{table}[t]
  \centering
  \renewcommand{\arraystretch}{1.2}
  \setlength{\tabcolsep}{7pt}
  \caption{BD-mAP gains of JRD-based VCM on different JRD models, tasks, and base codecs[Unit:\%].}
    \begin{tabular}{c|c|cccc}
    \toprule
    Base codec & Method & OD & IS & KPD & Avg. \\
    \midrule
    \multirow{7}[2]{*}{VVC} & EL-JRD & 2.058 & 2.346 & 2.823 & 2.409 \\
          & BC-JRD & 2.589 & 2.638 & 4.169 & 3.132 \\
          & DT-JRD & 2.694 & 2.233 & 3.783 & 2.903 \\
          & MT-JND & 0.883 & 1.310 & 3.117 & 1.770 \\
          & MJ-REC & 1.190 & 2.303 & 3.195 & 2.229 \\
          & AMT-JRD & \textbf{3.801} & \textbf{3.579} & \textbf{4.202} & \textbf{3.861} \\
          & GT-JRD & 5.515 & 5.951 & 4.917 & 5.461 \\
    \midrule
    \multirow{7}[2]{*}{JPEG} & EL-JRD & 5.065 & 5.486 & 10.337 & 6.963 \\
          & BC-JRD & 4.891 & 6.646 & 11.006 & 7.515 \\
          & DT-JRD & 5.791 & \textbf{6.866} & 10.241 & 7.633 \\
          & MT-JND & 4.376 & 4.681 & 10.895 & 6.651 \\
          & MJ-REC & 5.027 & 5.818 & 10.502 & 7.116 \\
          & AMT-JRD & \textbf{5.848} & 6.674 & \textbf{11.137} & \textbf{7.886} \\
          & GT-JRD & 6.626 & 7.555 & 11.156 & 8.446 \\
    \bottomrule
    \end{tabular}%
  \label{BD-mAP}%
\end{table}%

As shown in Fig. \ref{VVC}, the VVC all-intra method adopts a uniform quantization strategy without region-level perceptual control, leading to the lowest accuracy among all methods. In contrast, GT-JRD represents an ideal upper bound, in which the quantization is driven by ground truth JRD maps. Due to the nonlinear response of machine vision models to traditional codecs such as VVC and JPEG, which are originally designed for the HVS, slight discontinuities can be observed in the rate-accuracy curves of GT-JRD and other JRD-based methods near the perceptual threshold, as previously revealed in our earlier work\cite{zhang2023learning,liu2024dtjrd}. With its precise and robust multi-task perceptual threshold modeling capability, the AMT-JRD-based VCM demonstrates consistently superior rate-accuracy performance across the three tasks of OD, IS, and KPD. A similar trend is observed in Fig. \ref{JPEG}. While the bitrates under JPEG are generally higher than those in VVC, the proposed AMT-JRD-based VCM maintains competitive performance with effective region-level perceptual control.

To quantify the overall improvement, Table \ref{BD-mAP} reports the Bjontegaard Delta mAP (BD-mAP) gain for each method. BD-mAP is used to measure the average improvement of accuracy in machine vision tasks at the same bit rate. With VVC as the base codec, AMT-JRD-based VCM achieves 3.801\%, 3.579\%, and 4.202\% BD-mAP gains in detection, segmentation, and keypoint tasks, respectively. In other words, the AMT-JRD-based VCM achieved an average bit rate reduction of 16.614\%, 15.662\%, and 32.288\% across the three tasks, respectively. When using JPEG as the base codec, the AMT-JRD-based VCM achieved even greater BD-mAP gains, with improvements of 5.848\%, 6.674\%, and 11.137\% across the three tasks, respectively. The AMT-JRD-based VCM achieved an average BD-mAP of 7.886\%, second only to the GT-JRD-based VCM, once again demonstrating its performance closest to the ideal benchmark. The coding results on both VVC and JPEG demonstrate that the proposed AMT-JRD-based VCM not only has strong consistency across multiple vision tasks but also has good generalization capability across codecs.

\begin{table}[t]
  \centering
  \caption{BD-mAP of JRD-based VCM on cross-task validation[Unit:\%].}
  \renewcommand{\arraystretch}{1.2}
    \begin{tabular}{c|ccc|ccc}
    \toprule
    \multirow{2}[4]{*}{Task\textbackslash{}Method} & \multicolumn{3}{c|}{AMT-JRD} & \multicolumn{3}{c}{GT-JRD} \\
    \cmidrule(lr){2-4} \cmidrule(lr){5-7}  & OD & IS & KPD & OD & IS & KPD \\
    \midrule
    OD   & \textbf{3.801} & 2.844 & 3.219 & \textbf{5.515} & 3.977 & 2.810 \\
    IS   & 3.088 & \textbf{3.579} & 3.574 & 4.815 & \textbf{5.951} & 3.665 \\
    KPD   & 3.715 & 3.580 & \textbf{4.202} & 4.047 & 4.096 & \textbf{4.917} \\
    \bottomrule
    \end{tabular}%
  \label{crosstask}%
\end{table}%

\begin{table}[t]
\centering
\caption{Comparison of JRD model on accuracy, size, and computational cost.}
\renewcommand{\arraystretch}{1.2}
\setlength{\tabcolsep}{2pt}
\begin{tabular}{ccccccc}
\toprule
Model & HVS & Multi-Task & Params (M) & FLOPs (G) & $E_A$ \\
\midrule
EL-JRD  & \ding{55} & \ding{55} & 484.200($\times$3) & 936.408($\times$3) & 5.775  \\
BC-JRD  & \ding{55} & \ding{55} & \underline{31.587}($\times$3) & 989.184($\times$3) & 4.386  \\
DT-JRD & \ding{55} & \ding{51} & 305.655 & \underline{15.264} & \underline{4.052} \\
MT-JND & \ding{51} & \ding{51} & 34.082 & 30.752 & 4.660 \\
MJ-REC & \ding{51} & \ding{51} & \textbf{11.085} & 58.076 & 4.884  \\
AMT-JRD (Ours) & \ding{55} & \ding{51}  & 77.374 & \textbf{9.934} & \textbf{3.781}\\
\bottomrule
\end{tabular}
\label{tab:model-comparison}
\end{table}


\begin{figure}[htbp]
	\centering
        \includegraphics[width=0.45\textwidth]{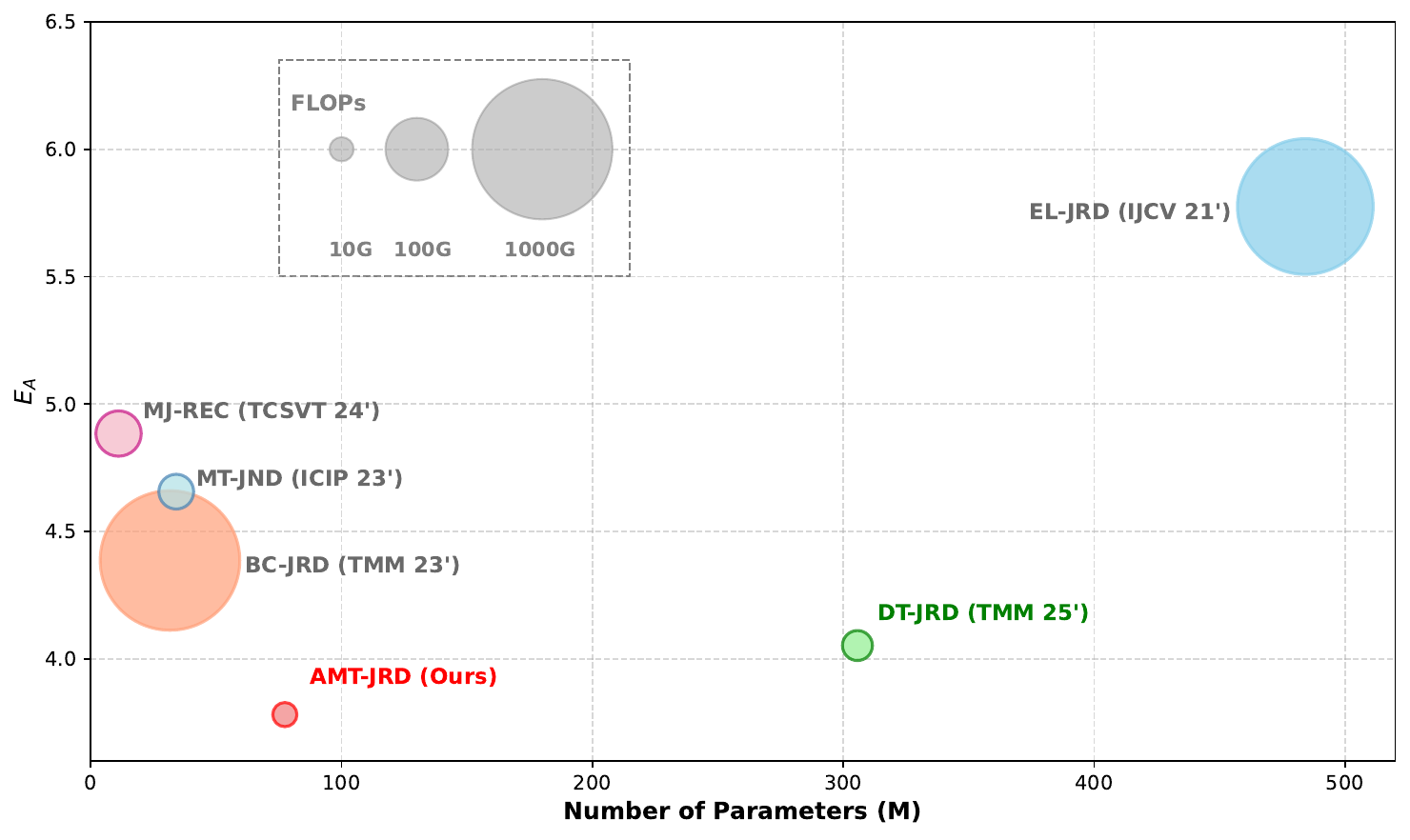}
        \captionsetup{justification=justified}
        \caption{Complexity and accuracy of JRD models on the MT-JRD dataset.}
        \label{Complexity}
\end{figure}

\begin{figure*}[t]
\captionsetup{justification=justified}
\centering
        \subfloat[]{
        \includegraphics[width=0.115\textwidth]{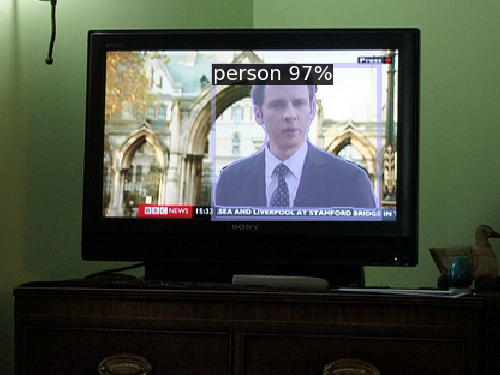}}
        \subfloat[]{
        \includegraphics[width=0.115\textwidth]{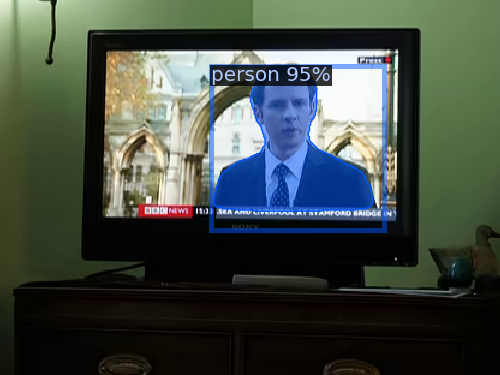}}
        \subfloat[]{
        \includegraphics[width=0.115\textwidth]{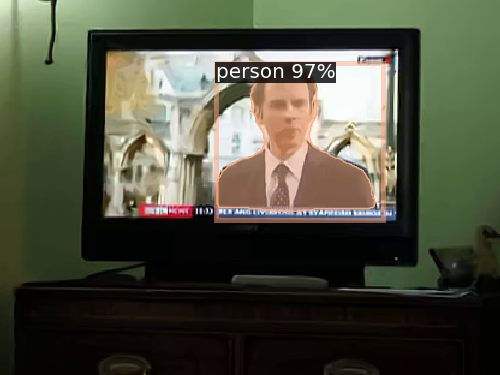}}
        \subfloat[]{
        \includegraphics[width=0.115\textwidth]{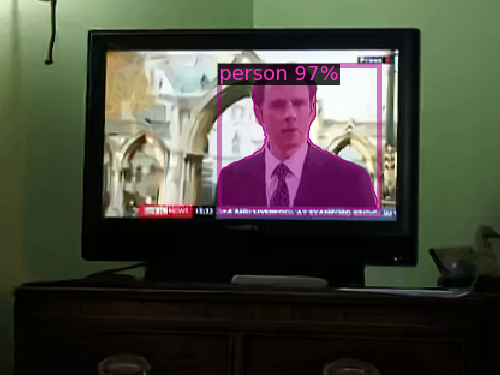}}
        \subfloat[]{
        \includegraphics[width=0.115\textwidth]{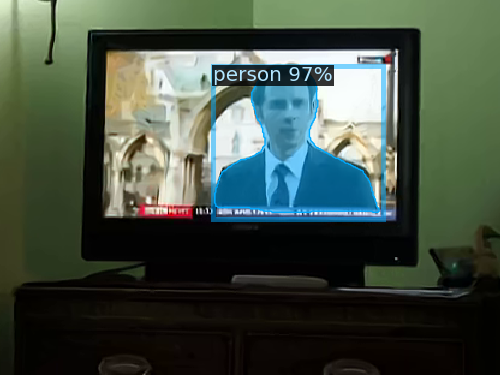}}
        \subfloat[]{
        \includegraphics[width=0.115\textwidth]{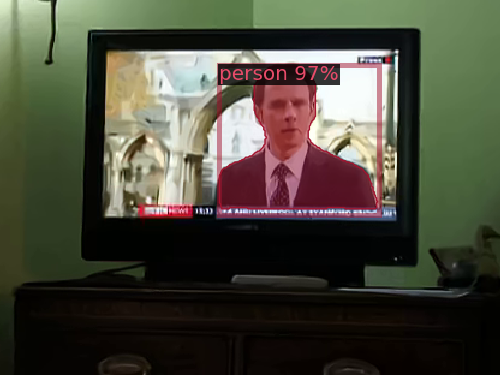}}
        \subfloat[]{
        \includegraphics[width=0.115\textwidth]{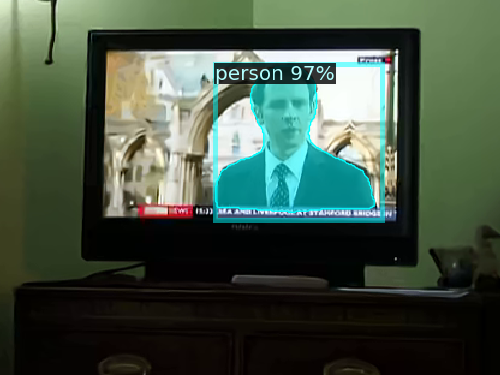}}
        \subfloat[]{
        \includegraphics[width=0.115\textwidth]{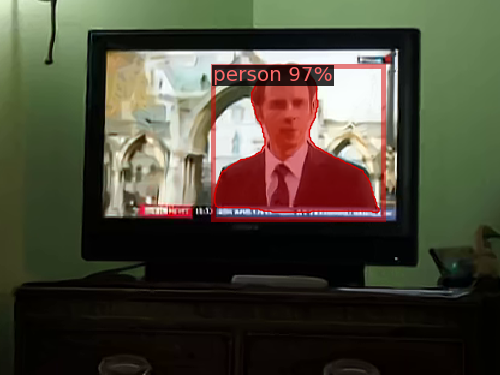}}
        \\[-2ex]
        \subfloat[]{
        \includegraphics[width=0.115\textwidth]{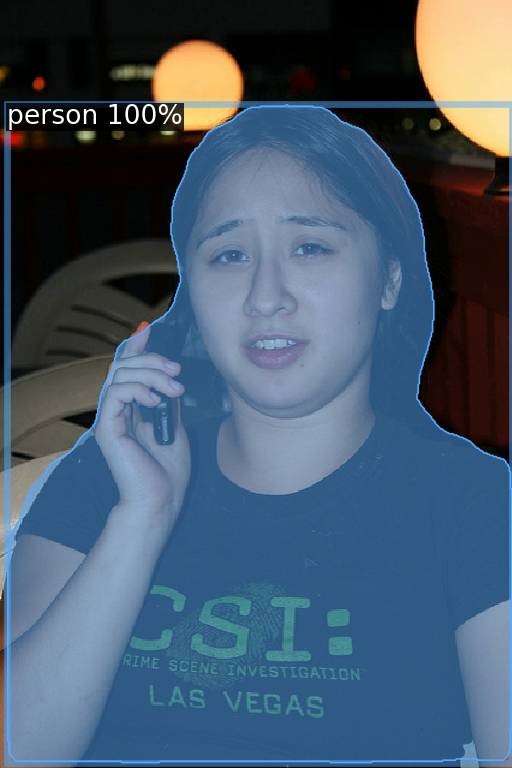}}
        \subfloat[]{
        \includegraphics[width=0.115\textwidth]{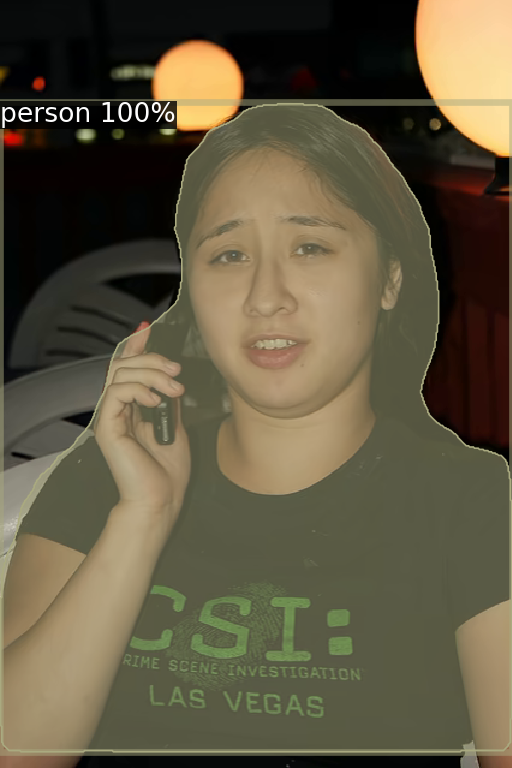}}
        \subfloat[]{
        \includegraphics[width=0.115\textwidth]{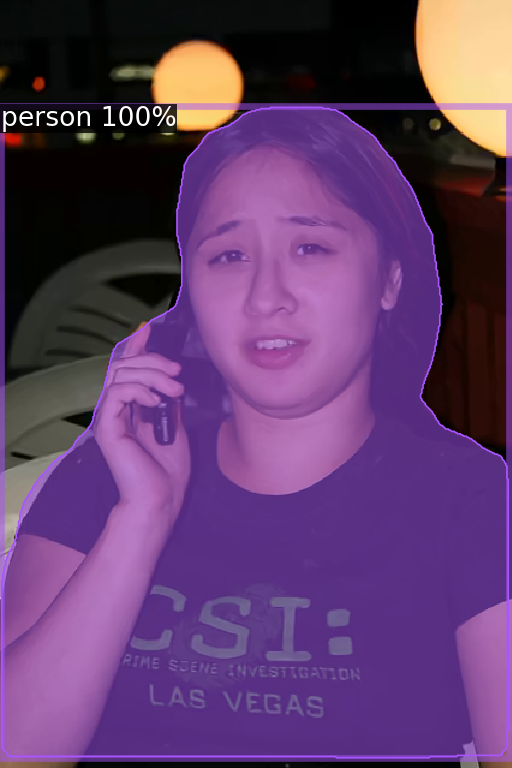}}
        \subfloat[]{
        \includegraphics[width=0.115\textwidth]{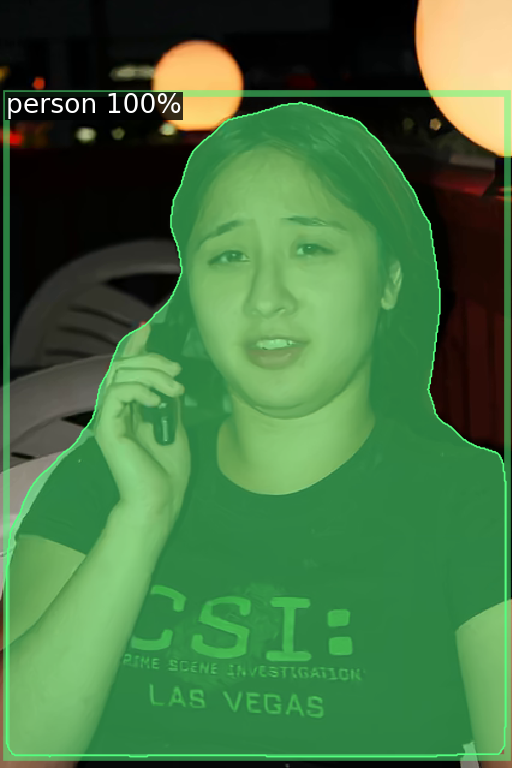}}
        \subfloat[]{
        \includegraphics[width=0.115\textwidth]{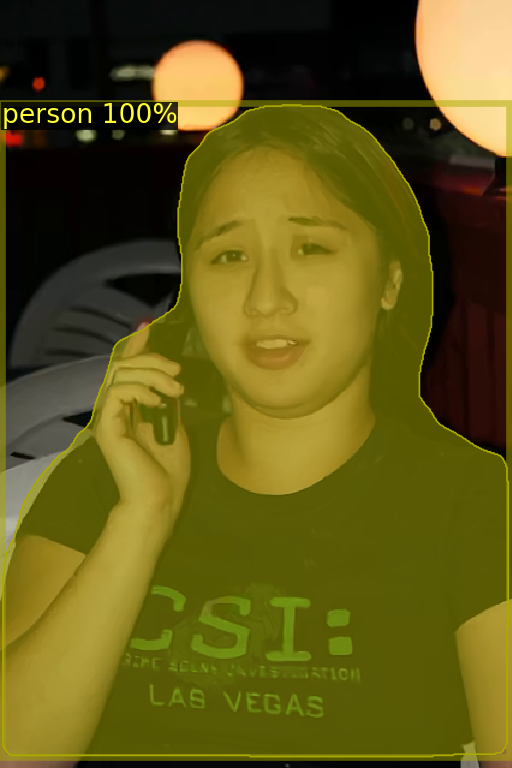}}
        \subfloat[]{
        \includegraphics[width=0.115\textwidth]{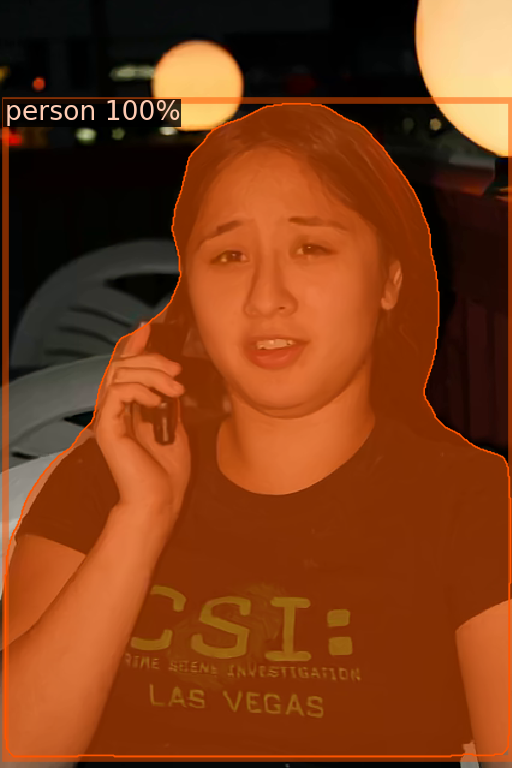}}
        \subfloat[]{
        \includegraphics[width=0.115\textwidth]{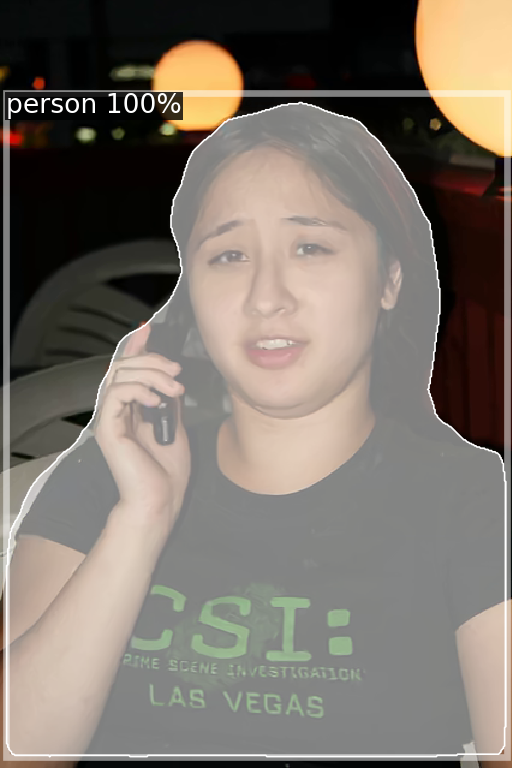}}
        \subfloat[]{
        \includegraphics[width=0.115\textwidth]{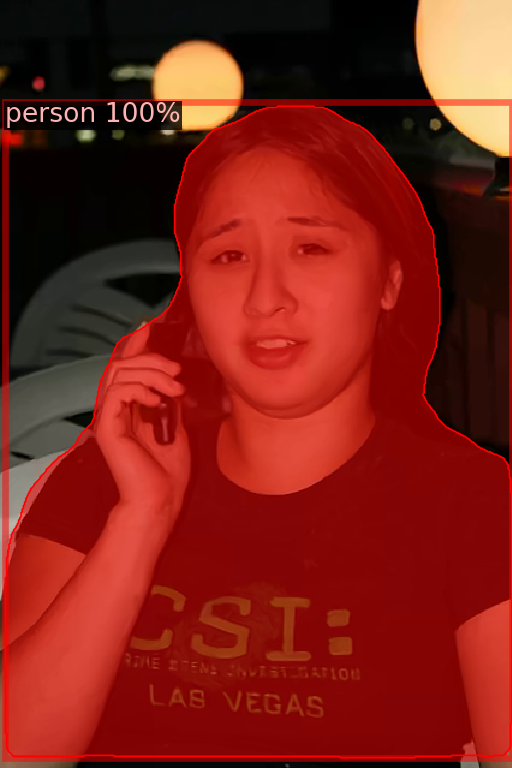}}
        \\[-2ex]
        \subfloat[]{
        \includegraphics[width=0.115\textwidth]{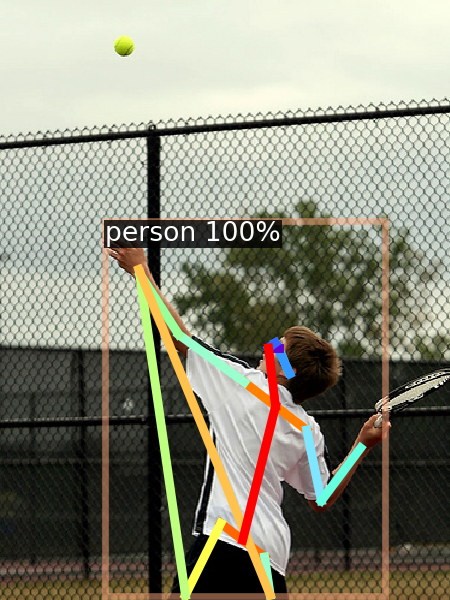}}
        \subfloat[]{
        \includegraphics[width=0.115\textwidth]{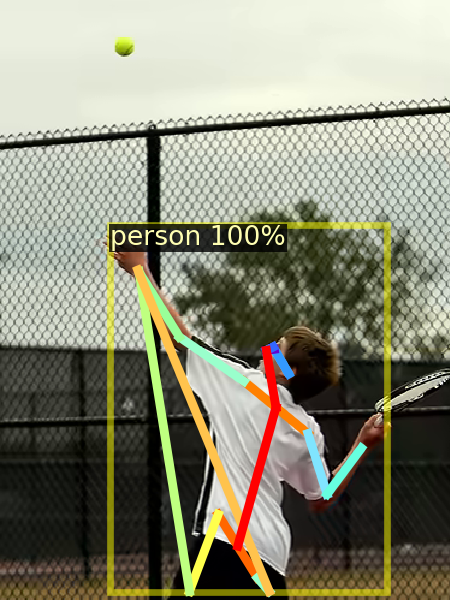}}
        \subfloat[]{
        \includegraphics[width=0.115\textwidth]{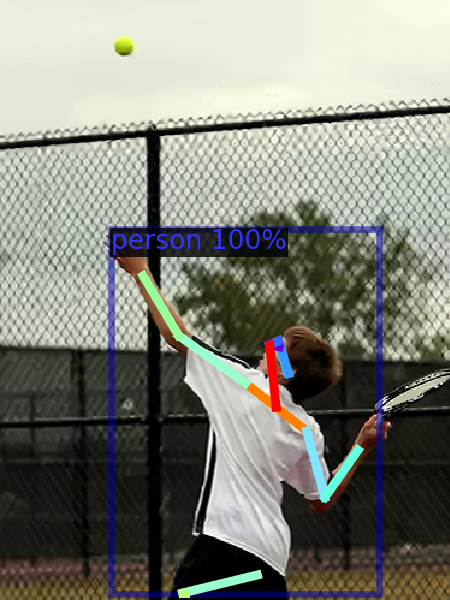}}
        \subfloat[]{
        \includegraphics[width=0.115\textwidth]{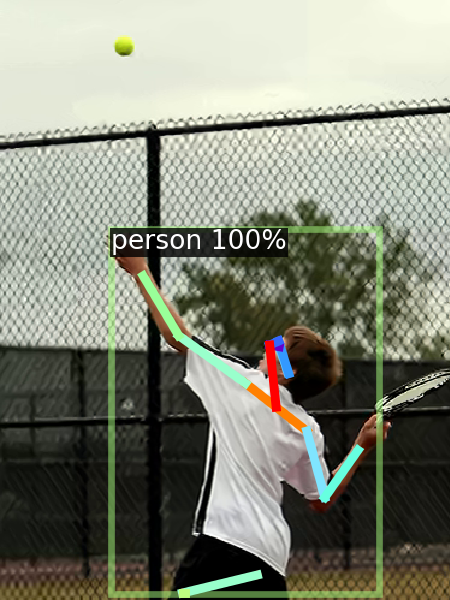}}
        \subfloat[]{
        \includegraphics[width=0.115\textwidth]{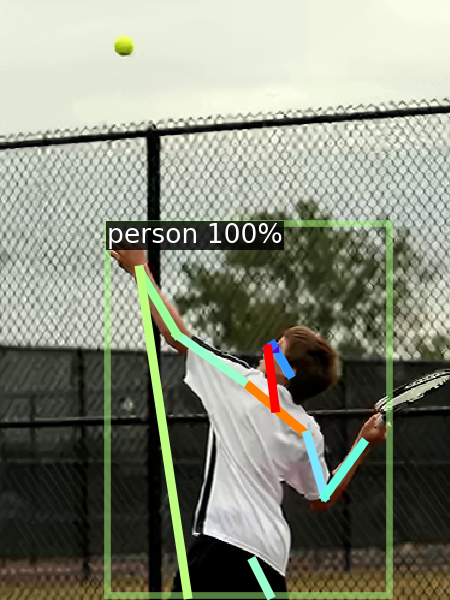}}
        \subfloat[]{
        \includegraphics[width=0.115\textwidth]{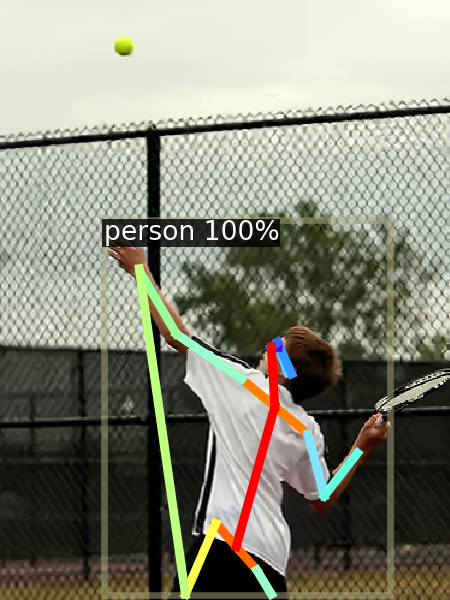}}
        \subfloat[]{
        \includegraphics[width=0.115\textwidth]{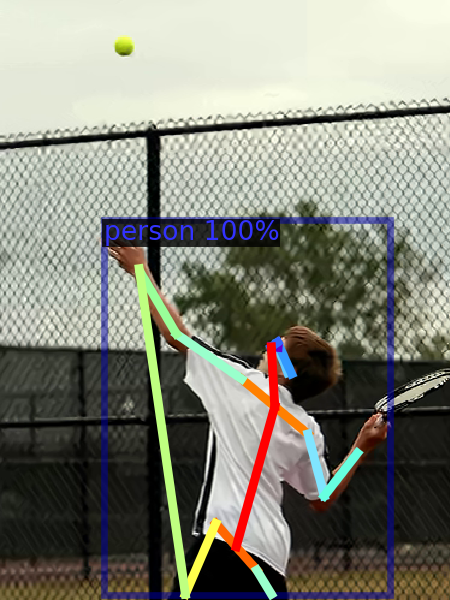}}
        \subfloat[]{
        \includegraphics[width=0.115\textwidth]{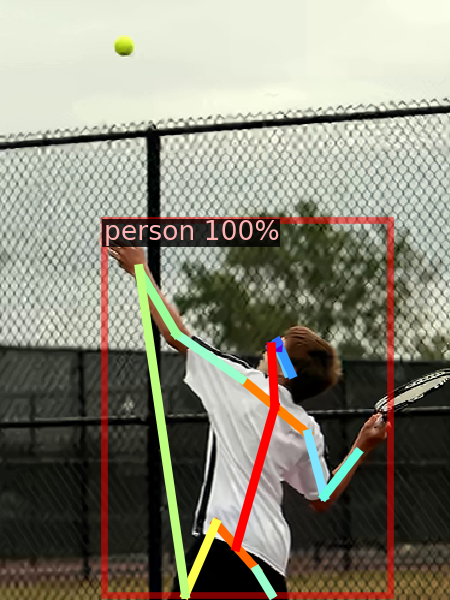}}
	\caption{Visualized machine analytical results of compressed images from different JRD-based coding methods. The first, second, and third lines correspond to OD, IS, and KPD, respectively. (a)(i)(q) original images, (b)(j)(r) VVC all-intra, (c)(k)(s) EL-JRD-based VCM, (d)(l)(t) BC-JRD-based VCM, (e)(m)(u) DT-JRD-based VCM, (f)(n)(v) MT-JND-based VCM, (g)(o)(w) MJ-REC-based VCM, (h)(p)(x) AMT-JRD-based VCM. The bit rate, confidence, and similarity metric of the detected objects are listed as triples. (b) \{0.2757, 0.9500, 0.8881\}, (c) \{0.2665, 0.9719, 0.9335\}, (d) \{0.2710, 0.9705, 0.9380\}, (e) \{0.2660, 0.9744, 0.9623\}, (f) \{0.2710, 0.9705, 0.9380\}, (g) \{0.2751, 0.9749, 0.9521\}, (h) \{0.2660, 0.9744, 0.9623\}, (j) \{0.2854, 0.9990, 0.9644\}, (k) \{0.2803, 0.9991, 0.9679\}, (l) \{0.2768, 0.9990, 0.9649\}, (m) \{0.2713, 0.9990, 0.9701\}, (n) \{0.2731, 0.9993, 0.9723\}, (o) \{0.2768, 0.9990, 0.9649\}, (p) \{0.2676, 0.9990, 0.9746\}, (r) \{0.4627, 0.9982, 0.9745\}, (s) \{0.4089, 0.9995, 0.7651\}, (t) \{0.4089, 0.9995, 0.7651\}, (u) \{0.3934, 0.9993, 0.9075\}, (v) \{0.4301, 0.9995, 0.9876\}, (w) \{0.4301, 0.9995, 0.9876\}, (x) \{0.4301, 0.9995, 0.9876\}}
	\label{visualization}
\end{figure*}

Additionally, we validated the cross-task JRD-based VCM method, with the BD-mAP gains recorded in Table \ref{crosstask}. VVC was employed as the base codec, while AMT-JRD and GT-JRD served as the predicted and ground-truth JRD values, respectively. On the one hand, encoding using predicted or ground-truth JRDs from other tasks also shows improvement compared to VVC. AMT-JRD leverages semantic similarity across three machine vision tasks through generalized feature extraction and achieves task-agnostic perceptual compression via the VCM method. On the other hand, the greatest performance gains are achieved when encoding using the JRD for this task. Taking the AMT-JRD model as an example, when using the predicted JRDs from the IS and KPD tasks for coding, the BD-mAP gains on the OD task are 2.844\% and 3.219\%, respectively, both lower than the 3.801\% gain achieved when using the OD-specific JRDs. Similar trends are also observed with GT-JRD and on other tasks. These results further validate that the specialized feature extraction in AMT-JRD effectively captures fine-grained semantic features critical to each task.

\subsection{Computational Complexity Analysis and Visualization}
We further analyze the number of parameters and computational complexity of each model, as illustrated in Table \ref{tab:model-comparison} and Fig. \ref{Complexity}. AMT-JRD achieves the lowest prediction error with an $E_A$ of 3.781, significantly outperforming all baseline methods. Meanwhile, it maintains an extremely low computational complexity with only 9.934G FLOPs, which is approximately 1\% of that required by repeated binary-classification models such as EL-JRD and BC-JRD. Although the model has a moderate number of parameters (77.374M), it strikes a favorable balance between performance and efficiency. Compared to other HVS-aware models (e.g., MJ-REC), AMT-JRD enables multi-task prediction without introducing additional architectural components, demonstrating the effectiveness of its core design. These advantages render AMT-JRD a scalable and practical solution for real-world machine vision applications.

Finally, we present a visual comparison of AMT-JRD-based VCM against other methods across multiple machine vision tasks in Fig. \ref{visualization}. Compared to VVC all-intra and other JRD-based VCM approaches, the AMT-JRD-based VCM achieves a superior rate-accuracy trade-off by preserving critical semantic details for machine vision through precise JRD modeling.

\section{conclusions}
\label{section:conclusion}
In this paper, we propose a multi-task JRD (MT-JRD) dataset and an
Attribute-assisted MT-JRD (AMT-JRD) model for Video Coding
for Machines (VCM). First, we created an MT-JRD dataset for object detection, instance segmentation, and keypoint detection, and conducted a detailed analysis of the commonalities and differences in JRD distributions across different tasks. Second, inspired by multi-task model architecture design and perceptual factors, we propose an AMT-JRD prediction model. Through the effective integration of the Generalized Feature Extraction Module (GFEM), the Specialized Feature Extraction Module (SFEM), and the Attribute Feature Fusion Module (AFFM), AMT-JRD not only balances the capabilities of task-general and task-specific feature extraction but also compensates for object attribute information to enhance perception threshold modeling. Finally, we implement the JRD-based VCM using VVC and JPEG and achieve a significant coding gain. The proposed MT-JRD coding framework enhances VCM coding efficiency and multi-task generalization.

\bibliographystyle{IEEEtran}
\bibliography{IEEEabrv,reference}

\begin{IEEEbiography}[{\includegraphics[width=1in,height=1.25in,clip,keepaspectratio]{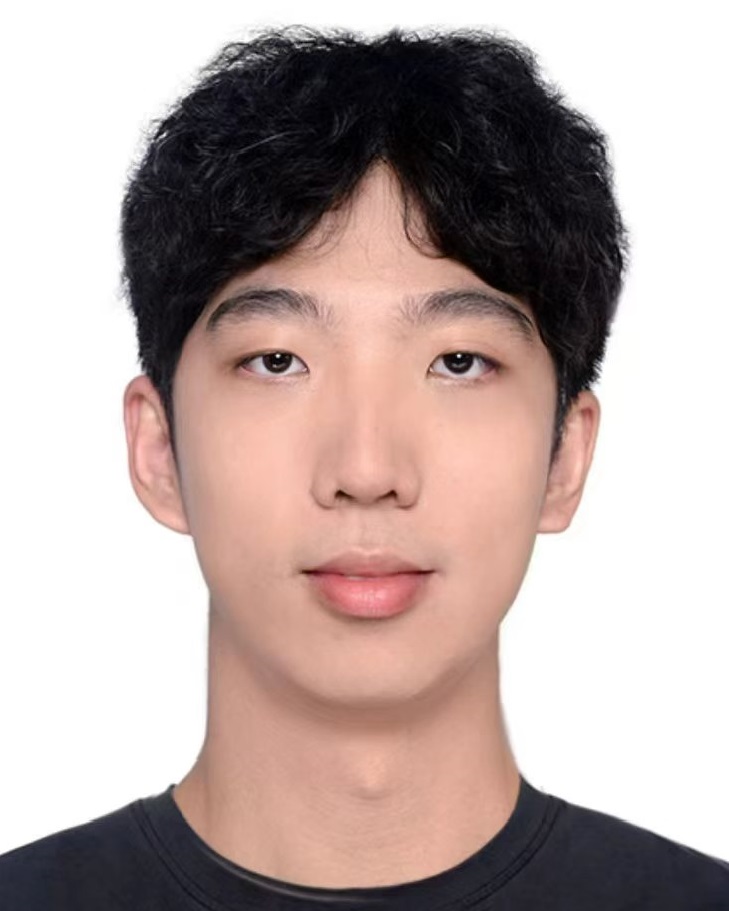}}]{Junqi Liu}
received the B.E. degree in electronic information science and technology from Sun Yat-sen University, China, in 2024. He is currently pursuing the M.E. degree with the School of Electronics and Communication Engineering, Sun Yat-sen University, Shenzhen, China. His current research interests include image/video coding for machines, quality assessment, and deep learning.
\end{IEEEbiography}

\begin{IEEEbiography}[{\includegraphics[width=1in,height=1.25in,clip,keepaspectratio]{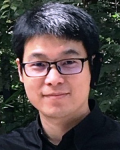}}]{Yun Zhang}
(Senior Member, IEEE) received the B.S. and M.S. degrees in electrical engineering from Ningbo University, Ningbo, China, in 2004 and 2007, respectively, and the Ph.D. degree in computer science from the Institute of Computing Technology, Chinese Academy of Sciences (CAS), Beijing, China, in 2010. From 2009 to 2014, he was a Visiting Scholar with the Department of Computer Science, City University of Hong Kong, Kowloon, Hong Kong. From 2010 to 2022, he was a Professor/Associate Professor with the Shenzhen Institutes of Advanced Technology, CAS, Shenzhen, China. Since 2022, he has been a Professor with the School of Electronics and Communication Engineering, Sun Yat-sen University (Shenzhen Campus), Guangdong, China. His research interests mainly include 3D visual signal processing, image/video compression, visual perception, and machine learning.
\end{IEEEbiography}

\begin{IEEEbiography}[{\includegraphics[width=1in,height=1.25in,clip,keepaspectratio]{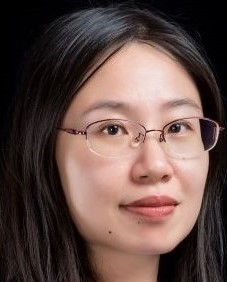}}]{Xiaoxia Huang}
(Senior Member, IEEE) received the B.E. and M.E. degrees in electrical engineering from the Huazhong University of Science and Technology, Wuhan, China, in 2000 and 2002, respectively, and the Ph.D. degree in electrical and computer engineering from the University of Florida, Gainesville, FL, USA, in 2007. She is currently a Professor with the School of Electronics and Communication Engineering, Sun Yat-sen University. Her research interests include green wireless communication networks, intelligent wireless networks, and wireless computing.
\end{IEEEbiography}

\begin{IEEEbiography}[{\includegraphics[width=1in,height=1.25in,clip]{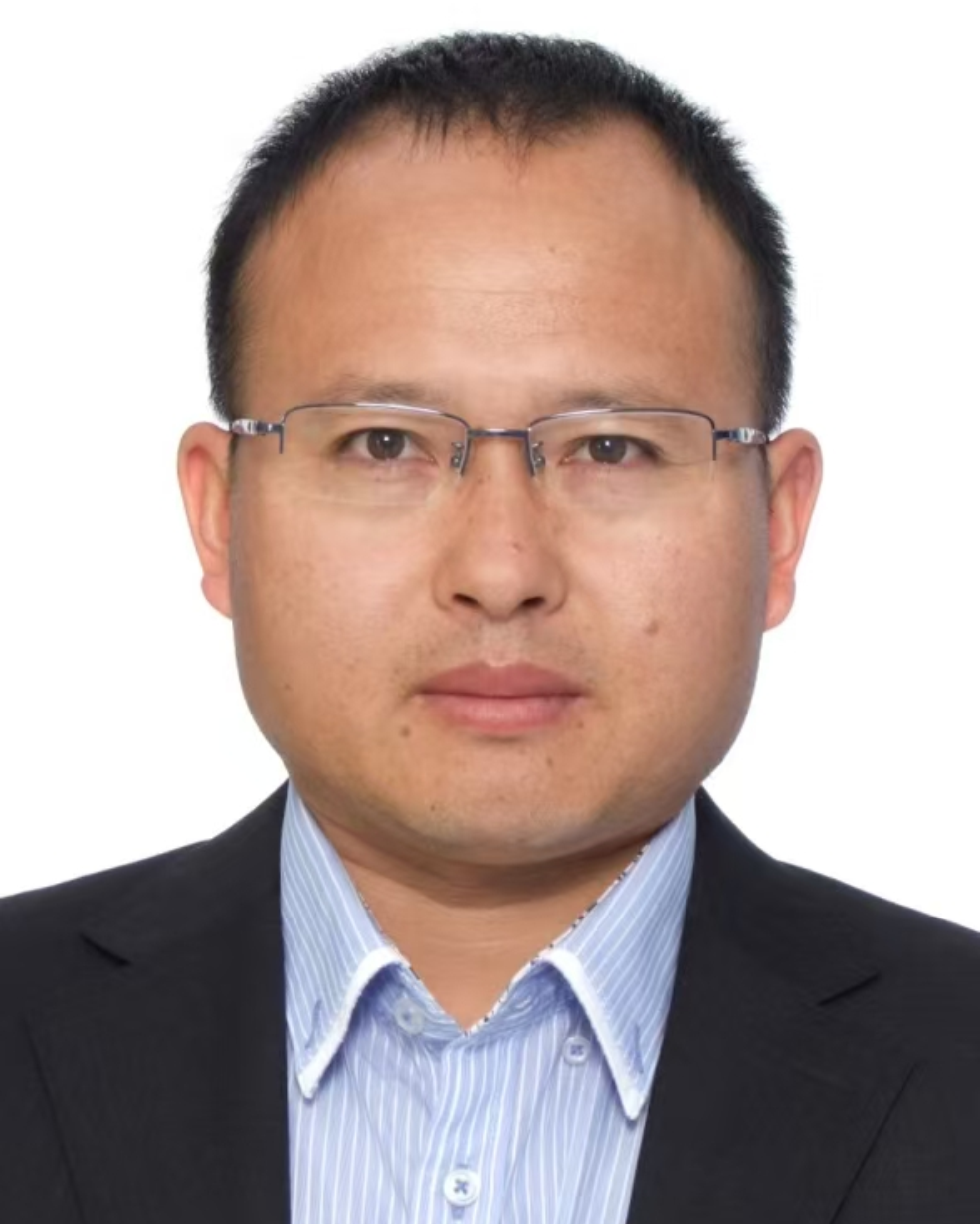}}]{Long Xu}
(Member, IEEE) received the M.S. degree in applied mathematics from Xidian University, Xi’an, China, in 2002, and the Ph.D. degree from the Institute of Computing Technology, Chinese Academy of Sciences (CAS), Beijing,
China. He was a Post-Doctoral Researcher with the Department of Computer Science, City University of Hong Kong, from July 2009 to December 2012, and the Department of Electronic Engineering, The Chinese University of Hong Kong, from August 2009 to December 2012. From January 2013 to March 2014, he was a Post-Doctoral Researcher with the School of Computer Engineering, Nanyang Technological University, Singapore. Currently, he is
with the Faculty of Electrical Engineering and Computer Science, Ningbo University, as a Full Professor. His current research interests include image/video processing, solar radio astronomy, wavelet, machine learning, and computer vision. He was selected into the 100-Talents Plan, CAS, in 2014.
\end{IEEEbiography}

\begin{IEEEbiography}[{\includegraphics[width=1in,height=1.25in,clip,keepaspectratio]{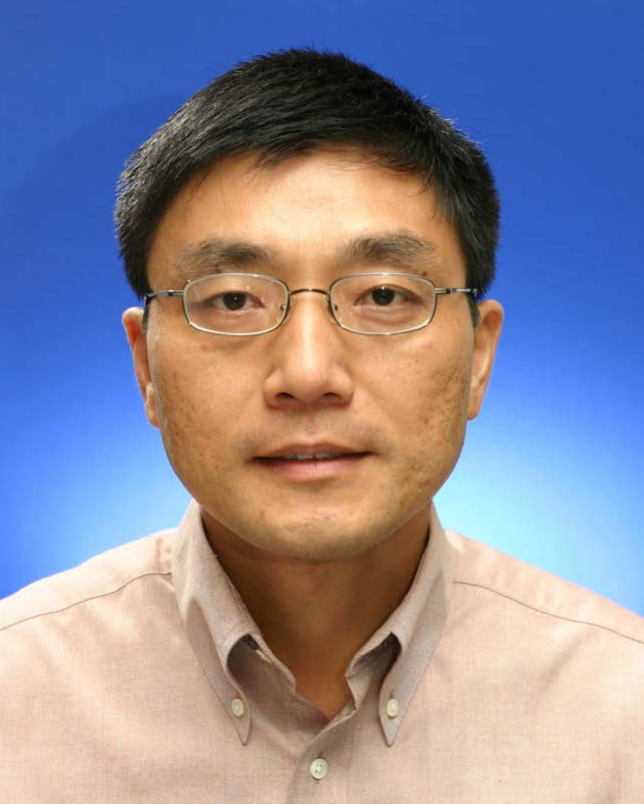}}]{Weisi Lin}
(Fellow, IEEE) received the Ph.D. degree from the King’s College London, U.K. He is currently a Professor with the College of Computing and Data Science, Nanyang Technological University. His research interests include image processing, perceptual signal modeling, video compression, and multimedia communication, in which he has published over 200 journal articles, over 230 conference papers, filed seven patents, and authored two books. He is a fellow of IET and an Honorary Fellow of
Singapore Institute of Engineering Technologists. He was the Technical Program Chair of IEEE ICME 2013, PCM 2012, QoMEX 2014, and IEEE VCIP 2017. He has been an invited/panelist/keynote/tutorial speaker at over 20 international conferences. He has been an Associate Editor of IEEE TRANSACTIONS ON IMAGE PROCESSING, IEEE TRANSACTIONS ON CIRCUITS AND SYSTEMS FOR VIDEO TECHNOLOGY, IEEE TRANSACTIONS ON MULTIMEDIA, and IEEE SIGNAL PROCESSING LETTERS. He was a Distinguished Lecturer of the IEEE Circuits and Systems Society from 2016 to 2017 and the Asia-Pacific Signal and Information Processing Association (APSIPA) from 2012 to 2013.
\end{IEEEbiography}

\vfill
\end{document}